\documentclass[
]{ceurart}

\usepackage{enumitem}

\usepackage{listings}
\begin{document}

\copyrightyear{2025}
\copyrightclause{Copyright for this paper by its authors.
  Use permitted under Creative Commons License Attribution 4.0
  International (CC BY 4.0).}

\conference{Joint Proceedings of the ACM IUI Workshops 2025, March 24-27, 2025, Cagliari, Italy}

\title{Can Generative AI Support Patients' \& Caregivers' Informational Needs? Towards Task-Centric Evaluation Of AI Systems}


\author[1]{Shreya Rajagopal}[email=shreyara@umich.edu]
\address[1]{University of Michigan, Ann Arbor}

\author[2]{Jae Ho Sohn}[email=JaeHo.Sohn@ucsf.edu]
\address[2]{University of California, San Francisco}

\author[3]{Hari Subramonyam}[email=harihars@stanford.edu]
\address[3]{Stanford University, Stanford}

\author[4]{Shiwali Mohan}[email=shiwali.mohan@gmail.com]
\address[4]{SRI International}
\cormark[1]

\cortext[1]{Corresponding author.}

\begin{abstract}
  Generative AI systems such as ChatGPT and Claude are built upon language models that are typically evaluated for accuracy on curated benchmark datasets. Such evaluation paradigms measure predictive and reasoning capabilities of language models but do not assess if they can provide information that is useful to people. In this paper, we take some initial steps in developing an evaluation paradigm that centers human understanding and decision-making. We study the utility of generative AI systems in supporting people in a concrete task - making sense of clinical reports and imagery in order to make a clinical decision. We conducted a formative need-finding study in which participants discussed chest computed tomography (CT) scans and associated radiology reports of a fictitious close relative with a cardiothoracic radiologist. Using thematic analysis of the conversation between participants and medical experts, we identified commonly occurring themes across interactions, including clarifying medical terminology, locating the problems mentioned in the report in the scanned image, understanding disease prognosis, discussing the next diagnostic steps, and comparing treatment options. Based on these themes, we evaluated two state-of-the-art generative AI systems against the radiologist's responses. Our results reveal variability in the quality of responses generated by the models across various themes. We highlight the importance of patient-facing generative AI systems to accommodate a diverse range of conversational themes, catering to the real-world informational needs of patients. 
\end{abstract}


\maketitle
\section{Introduction}
Conversational systems built upon generative AI technology are general-purpose tools that enable people to access information through natural conversations ~\cite{dwivedi2023so}. Healthcare is primed to have significant benefits with reliable, robust and trustworthy adoption of such systems, assisting different medical interactions/conversations involving patients, doctors, clinicians, and healthcare administrators. Purportedly, generative AI systems have medical commonsense at par with doctors as evidenced by their high scores in standard medical knowledge evaluation paradigms like the United States Medical Licensing Examination (USMLE)~\cite{jin2021disease}. Despite impressive results on benchmark medical and clinical datasets, the case for their adoption in real world clinical scenarios is still unclear. While several clinical researchers and practitioners have expressed excitement about their potential in healthcare \cite{cascella2023evaluating, clusmann2023future}, several critical questions must be answered before their adoption can begin. These questions include what role generative AI systems can play in a healthcare system, how their behavior should be modulated, and how their performance can be evaluated. Answering these questions requires going beyond benchmark datasets prevalent in AI \& ML research to understand the nature of real world problems and characteristics of desired solutions.     

In this paper, we study a real world clinical usecase for generative AI systems. Patients (along with their caregivers) who are better educated about their disease have better outcomes, especially in serious illnesses such as cancer ~\cite{adams2010improving, berkman2011, howell2017self, christiansen2023systematic}. However, they face complex information and learning challenges when navigating healthcare systems. They not only must learn what clinical terminology means but also what their medical reports convey, how the severity of their disease is diagnosed, what their treatment options are etc. This often leads to anxiety and stress not only from the disease but also from the complexity of understanding and managing it. To alleviate these challenges, healthcare systems have invested significantly in patient education resources. Meaningful and empathetic interactions with healthcare providers has a significant, positive impact on a patient's journey through the healthcare system \cite{gilligan2018patient}. While arguably the most useful, it also takes the experts' time away from their other critical responsibilities, adding to their time burden. On the other hand, patients often desire additional informational support and augment their understanding with generic and unverified information from search engines \cite{de2019quality} and forums \cite{alarifi2021understanding}. Integrating generative AI systems in patient-facing tools requires that generative AI systems be designed to teach and explain diagnosis, and prognosis; weigh risks, benefits, and costs of treatments; and understand complex human emotional and social contexts. Lack of adequate considerations can lead to harmful consequences for patients, including misinterpretation of generated output, over-reliance on AI, and poor disease management. 

We examine the scenario where patients and caregivers attempt to understand their medical scans and reports by interacting with a clinical expert. In collaboration with a cardiothoracic radiologist at a major university hospital (and a co-author), we conducted a formative need-finding study with $5$ participants. The study was framed as a conversational interview in which participants discussed chest computed tomography (CT) scans and associated radiology reports of a fictitious close relative with a radiologist. We limited the participants to a small number because the subject of the study is highly sensitive and could trigger anxiety in our participants. Additionally, each interview required $~3hrs$ of the radiologist's time. We analyzed the recorded interviews with inductive thematic analysis to identify emerging themes and categories in the conversations. Next, we evaluated two state-of-art multi-modal generative AI systems ChatGPT-4o~\cite{yang2023dawn} and Claude 3.5 Sonnet \cite{claude2024} - against the radiologist's responses in the realistic interview task. Our findings highlight the strengths and weaknesses of generative AI models in addressing the information needs of patients and caregivers.

Our work situates and evaluates generative AI capabilities in a specific human context of seeking information to enhance understanding. This context enables us to identify the variety of information patients seek and study if AI systems are up to the challenge. Further, generative AI systems are typically evaluated on datasets curated by and for clinicians. These datasets contain questions and answers that are most interesting to clinicians, while performance measurement includes criteria such as diagnostic accuracy. Such an approach introduces an implicit bias in the design and analysis of these systems - they are designed to replace clinicians by mimicking the output of their problem-solving process, following the \emph{AI-is-automation} philosophy. Instead, our research takes the \emph{AI-is-augmentation} view and centers patients and caregivers. We explore if generative AI systems can support humans - patients and caregivers - in a specific task - making sense of the their reports and scans and accessing knowledge about their disease and treatment.

Advancing the science of human-centered, task-centered evaluation of generative AI systems, our paper makes the following contributions. It, 
\begin{itemize}
    \item Identifies an opportunity for adoption of generative AI system in clinical practice and specifies an AI task reflecting a real world problem; 
    \item Characterizes the variability in types of information people seek when making sense of medical reports and scans;
    \item Proposes an evaluation approach for conversational systems quantitatively using semantic similarity as well as qualitatively, building upon the Gricean maxims of effective communication; 
    \item Evaluates two state-of-art generative AI systems - ChatGPT-4o and Claude 3.5 Sonnet - on the proposed evaluation rubric.
\end{itemize}
\section{Background}
\label{section_background}

Recently, large language models (LLMs) like PaLM~\cite{chowdhery2023palm}, GPTs~\cite{radford2019language, brown2020language, openai2023gpt4}, LLaMA~\cite{touvron2023llama, touvron2023llama2}, and Claude~\cite{claude2024} have significantly advanced the state-of-the-art in various natural language processing (NLP) tasks~\cite{zhao2023survey, yang2023harnessing}. This has inspired the development of many medical LLMs~\cite{singhal2023large, singhal2023towards, han2023medalpaca, toma2023clinical, yunxiang2023chatdoctor, liu2023medical} to assist the healthcare professionals and improve patient care~\cite{arora2023promise, thirunavukarasu2023large, patel2023chatgpt}. Based on publicly available LLMs, specialized medical versions of the models like ChatDoctor~\cite{yunxiang2023chatdoctor}, MedAlpacha~\cite{han2023medalpaca}, PMC-LLaMA~\cite{wu2023pmc}, BenTsao~\cite{wang2023huatuo}, and Clinical Camel~\cite{toma2023clinical} have come up in the past years. Models like Med-PaLM~\cite{singhal2023large} and Med-PaLM-2~\cite{singhal2023towards} built on top of PaLM have achieved close to human expert scores in the United States Medical Licensing Examination (USMLE)~\cite{jin2021disease}. Apart for only text-based models several multimodal models~\cite{liu2023visual, awadalla2023openflamingo, li2023blip, yang2023dawn} have come to prominence in the last year with the capability of processing both images, text and videos to generate responses. Similar to NLP  models such multimodal models have promising application in the medical domain. They can be used as visual question answering systems for patient medical reports, medical report summarization and analysis tasks among many other possibilities. In this work, we evaluate Claude~\cite{claude2024} and GPT-4V~\cite{yang2023dawn} in a patient-facing, clinical usecase. 

GPT-4 model with vision capabilities (GPT-4V)~\cite{yang2023dawn} has been identified as the most advanced LLM thus far, particularly in its application to radiology domain, where it has been compared against the current state-of-the-art (SOTA) models \cite{liu2023exploring}. The research, which referenced various sources, thoroughly examined GPT-4V across a broad spectrum of standard radiology text-based tasks, including MS-CXR-T\cite{bannur2023mscxr} and CheXbert\cite{smit2020chexbert}. The findings revealed that GPT-4 either surpasses or matches the performance of specialized fine-tuned radiology models. A detailed investigation \cite{yang2023dawn} of GPT-4V studied its utility in generating radiology reports. This investigation demonstrated that GPT-4V could accurately diagnose and recommend management based on X-ray images. The findings from this research indicate that GPT-4V holds significant promise as an AI assistant in radiology report generation. Claude \cite{claude2024} models are more recent but boast multi-modal reasoning capabilities at par to GPT and were the top-performing models on a recent evaluation on medical, multi-modal answer generation benchmarks \cite{toma-etal-2024-wanglab-mediqa}.

Despite the growing interest in the application of Generative AI models in the medical and healthcare domains, there are a few key issues that need to be addressed for their reliable and effective usage in such a critical domain. Several recent surveys~\cite{zhao2023survey, he2023survey,xie2023faithful} on the application of LLMs for medicine highlight the best practices and challenges involved in their application and development. Effective evaluation of AI systems build around LLMs is one such challenge. Particularly in the medical domain, the current benchmarks and metrics often fail to evaluate LLMs' overall capabilities and emerging abilities. Current benchmarks such as USMLE MedQA~\cite{jin2021disease} and MedMCQA~\cite{pal2022medmcqa} cover broad range of question-answering tasks but lack the information to evaluate on metrics like trustworthiness and helpfulness to different parties involved like patients, doctors and administrators~\cite{xie2023faithful}. It is imperative to prepare more domain and task specific evaluation protocols to adjudicate the properties of GenAI models. Singhal et al.~\cite{singhal2023large} created a benchmark comprising the most commonly searched health queries to evaluate LLM responses. TruthfulQA~\cite{lin2021truthfulqa} and HaluEval~\cite{li2023halueval} attempt to evaluate truthfulness and factual accuracy, but not for any specific medical domain. The analyses reported in our work is the first attempt to reflect upon the conversational patterns in real-world patient-doctor interactions and evaluate strengths and weakness of generative AI system responses against those of human experts.

\section{Preliminaries}
\label{sec:preliminaries}
\begin{figure}[ht] 
    \centering
    \includegraphics[width=0.9\linewidth]{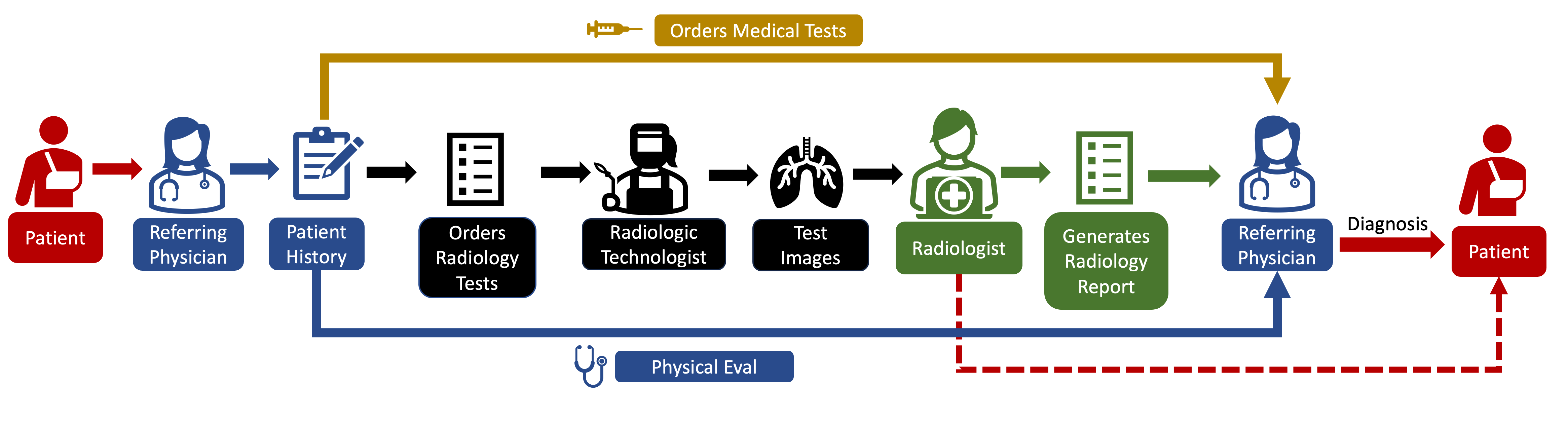}
    \caption{Patient journey through the medical system highlight various interactions they have with different healthcare experts.}
    \label{fig:patient-journey}
\end{figure}

The average patient's journey through the medical system in the United States is shown in Figure~\ref{fig:patient-journey}. Typically, a patient first enters the medical system and meets with a primary care physician (PCP) who collects their history, performs a physical evaluation, and orders further medical and radiology-based tests in order to inform a diagnosis. A radiologic technologist performs the radiology tests the referring PCP orders, and passes on these scans to the consulting radiologist. The radiologist analyses these scans and generates an associated radiology report. The referring physician receives these scans, reports, and the results of other medical tests they had ordered, and combines this information to arrive at an appropriate diagnosis. They then meet with the patient again to discuss the diagnosis, and followup and treatment options. It is relatively rare for radiologists to directly meet with their patients to help them understand their scans and radiology reports \cite{glazer2011invisible,kemp2017patient}.

A radiology report contains 3 major parts - Imaging techniques, Findings, and Impressions. Imaging techniques describe the specific imaging methodology used, and other parameters associated with the method. For example, in a chest CT, this section would describe the dimensional plane along with images that were collected (e.g. axial), the size of each voxel in the image, and whether or not intravenous contrast was administered. Findings describe whether or not any suspicious entities were found in the body parts scanned, and also describe their size, shape, and other appearance-based factors. Impressions are where the radiologists’ expertise most comes into play, with the radiologist giving their opinions on what the findings likely mean in the diagnostic context of the tests ordered by the patient’s physician. This could include likely diagnoses, the odds of each diagnosis, and suggested diagnostic follow-ups.

Since the radiology report is intended to aid the patient’s physician in arriving at a diagnosis, and not the patient themselves, the report contains medical terminology that is not accessible to the average patient. However, the patient receives immediate access to their radiology report once it is generated, in their electronic health records. Since it is atypical and oftentimes difficult for a patient to schedule a one-on-one consultation with their consulting radiologist to understand their scan and the associated report, patients have to wait two to three weeks on average to meet with their referring physician instead. This can tend to create anxiety in patients faced with medical information they cannot understand. Often, patients resort to unstructured, unverified resources such as Reddit for further information. 

An generative AI system can be useful during the period that the patient is waiting to meet with their PCP. The AI system can support a patient understand medical terminology used in the report and how the terms relate to the scan. It can help the patient understand the causes of their disease and its prognosis. It can support patients research various treatment options and weight their risks and benefits. Engaging with an intelligent information tool can prepare the patient (and their caregivers) for a productive meeting with their PCP. 


\section{Formative Study of Patient-Provider Interaction}
\label{sec:study}
We designed a formative study to closely resemble the medical context patients and caregivers find themselves in as they navigate the healthcare system (Section \ref{sec:preliminaries}) - the timepoint when they have access to their scans and reports and are preparing to meet with the PCP. We formulated the study as an interview between a participant simulating the role of a caregiver with a radiologist. It recreated the realistic interaction between a close relative of a patient receiving some worrying medical information, and the radiologist who developed the report. The study was conducted virtually on Zoom, with the experimenter and participant meeting first, and the radiologist joining at a later stage in the task. 

\subsection{Study}
\begin{figure}[ht]
    \centering
    \includegraphics[width=0.9\linewidth]{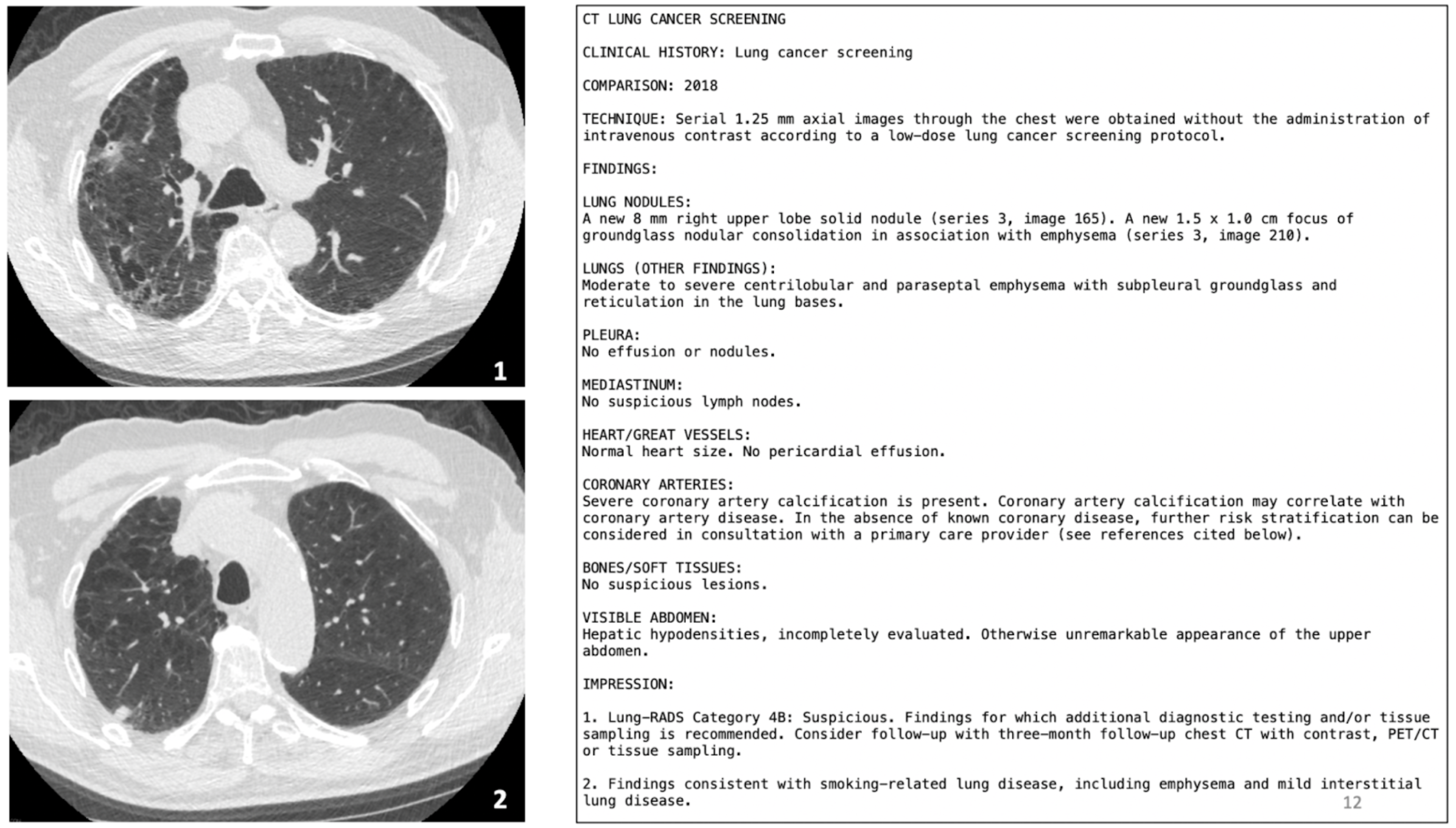}
    \caption{2 axial slices of the lung CT and the radiology report presented to study participants}
    \label{fig:case-report}
\end{figure}
\subsubsection{Materials} We created a case report (shown in Figure \ref{fig:case-report}) that comprised of $2$ axial slices of the lung CT and a corresponding radiology report with findings. The $2$ slices selected were the ones with evidence relevant for the findings in the report. The case report was developed using data collected for real cancer patients, modified and redacted for patient privacy purposes. The case was carefully selected such that it had some ambiguity both about visual evidence (i.e., there we two nodules that indicated malignancy) and the potential diagnosis (i.e., it wasn't certain that the paitent had cancer that this timepoint). These characteristics of the case enable us to understand how doctors communicate and manage uncertainty and how that might differ in Generative AI systems.

\subsubsection{Protocol} 
\begin{figure}[t]
    \centering
    \includegraphics[width=1\linewidth]{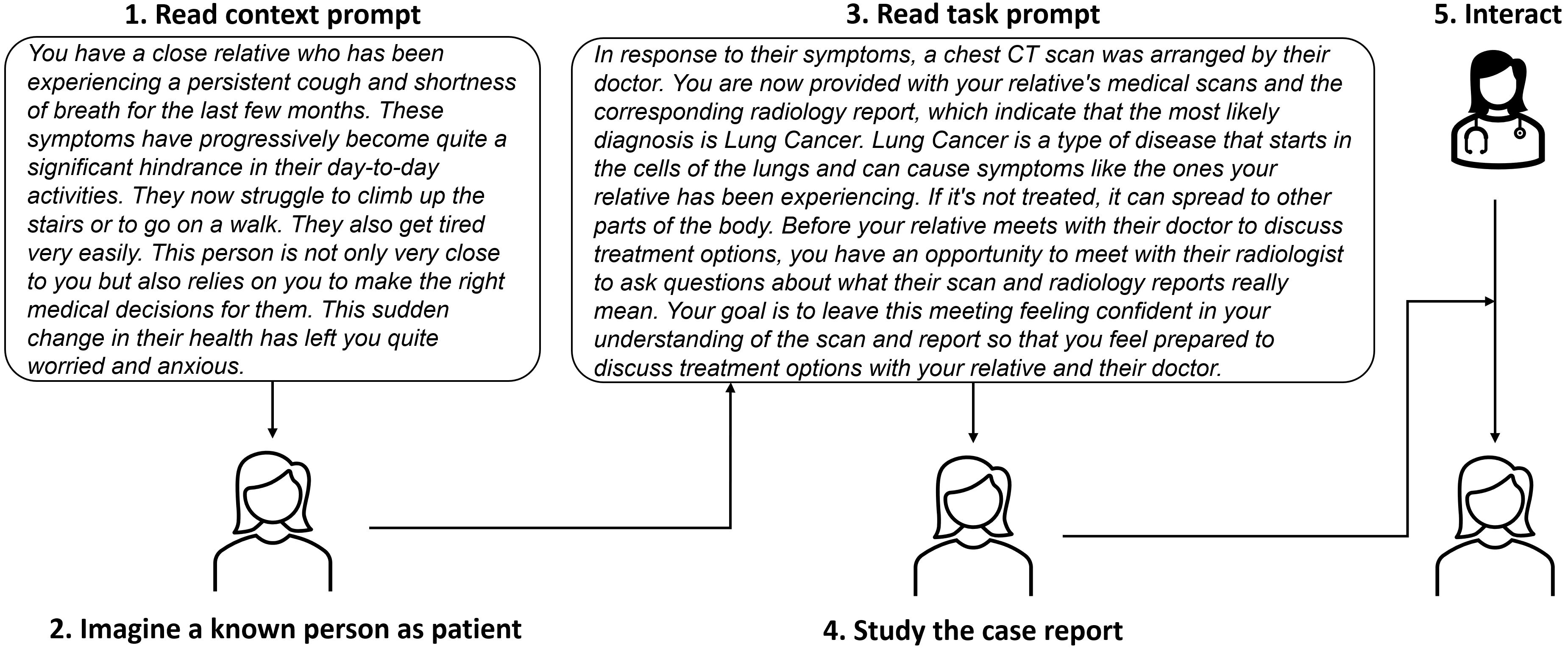}
    \caption{Protocol showing the context and task prompt as well as participant actions and interactions}
    \label{fig:protocol}
\end{figure}

Figure \ref{fig:protocol} shows the study protocol we implemented. The study progressed as follows (item numbers below correspond to stages of the study depicted in \ref{fig:protocol}):

\setlist{nolistsep}
\begin{enumerate}[noitemsep, left=0pt]
    \item Each participant was provided with the context prompt containing some symptoms their relative has been experiencing, and their role in their relative’s diagnostic journey.
    \item The participants were asked to imagine a close relative as patients and describe them in terms their age, sex, and relationship with the participant. They were also asked what other health conditions the imagined relative had simultaneously been dealing with. This was done to aid the follow-up interaction with the radiologist, where responses to certain concerns would be contingent on this information about patient demographics and comorbidities.
    \item The participants were provided with a task prompt that suggested that the most likely diagnosis was lung cancer, and described the participant’s goals in their upcoming interaction with their relative’s radiologist.
    \item The participants studied the case report (Figure \ref{fig:case-report}) shown on a screen and noted what they currently understood about their relative’s diagnosis, and the questions and concerns they would like to discuss with the radiologist in their upcoming interaction. they were also prompted to remember their emotional concerns regarding their relative’s well-being in addition to seeking information about the scans and reports. This step was designed to prepare the participant for meaningful interaction with the radiologist.
    \item The radiologist was then invited into the virtual room, and following initial introductions, the experimenter turned their microphone and camera off to allow for a natural interaction between the radiologist and the participant. The radiologist began by briefly discussing their role in the patient’s medical journey and introducing the patient to the problem at hand. The participants asked what questions came naturally to their mind as they looked at the case report. The scans and the report remained on the screen for the duration of the interaction, and the radiologist would often annotate the images on the screen to answer questions where referring to the scan was helpful.
\end{enumerate}

\subsubsection{Conducting the study} Our research company's internal review board conducted an \emph{Expedited} review of our study material and protocol and approved it with the risk determined to be \emph{minimal}. We recruited participants via an internal mailing list as well as through individual connections. During recruitment, we were careful to highlight the sensitive nature of the materials and the concern that participation in the study may trigger emotional response if the participants' close relative had gone through a similar medical experience. The email recruitment and the consent form prominently noted that the participant could request to stop the study at any point they desired without affecting their compensation. We recruited $5$ participants and each study session took close to $2$ hours. Each participant was paid $\$30$ as an Amazon gift card. The radiologist participating in the study is a from a major hospital at a public university and is a co-author. The interaction between the participant and the radiologist was recorded. 


\subsection{Inductive Thematic Analysis}
\label{sec:section_themes}
We transcribed the interaction sessions using Zoom's transcription service. Our final dataset consisted of $5$ documents averaging $3049$ ($R=1642-4614$) words. We analyzed this data set with inductive thematic analysis \cite{nowell2017thematic}. We started with a single interview and went through the transcript one sentence at a time, assigning codes that captured the content of the sentence, or a set of related sentences. For each new interview we analyzed, we used existing codes generated based on previous interviews and assigned new codes only when there were no existing codes to capture the content of a specific statement or set of statements. Each time a new set of codes was added based on an interview, we revisited previously coded interviews to check if the new codes could be assigned to any statements in them. We found that we reached data saturation at $5$ interviews with only $2$ new codes being added in analyzing the final interview. We identified a set of $91$ content-based codes across all transcribed participant-radiologist interactions, and we grouped these codes into a set of $10$ broad themes that best captured the relationship between the codes. 



The primary set of content-based codes were grouped together into the following $10$ themes. Each theme is described with some representative examples of quotations. These quotations consisting questions and answers are extracted and unmodified from the transcribed versions of the interview data, with the interviews these were drawn from indicated in parentheses. Quotations by the radiologist are indicated as (R-Interview Number).

\paragraph{\textbf{Theme 1}: Statistics of Lung Injury and Associated Damage to Other Organs} comprised of content including the likelihood of the damage found in the lungs for a patient with specific demographics, the correlation of this lung damage with damage to other organs, and the likelihood of damage to other organs as well. For example, \textit{ ``So usually, especially in this age group of my grandmother, emphysema can often lead to lung cancer, basically right?” (P03), ``I only have one question about specifically the coronary artery part where it says there is severe coronary artery calcification present. Do you know if that is related to lung cancer at all?” (P02), ``So it's very, very common for people, patients at age 80s to have coronary artery disease.” (R-P03), ``So now just having severe coronary artery calcification doesn't necessarily mean that your sister is going to have a heart attack, but it's certainly a statistical risk factor that we watch out for.” (R-P02)}

\paragraph{\textbf{Theme 2}: Role of Members of the Medical Team} included description of a typical patient's journey through the medical system, and how different members of the medical team had different roles in diagnosing and designing a treatment plan for the patient. This included subjects like the radiologist’s role, the role of the referring physician, the oncologist, and the multidisciplinary tumor board. E.g., \textit{``For diagnosis of cancer we have these multiple experts talking to each other and determining what is the best next course of action based on our combination of expertise.” (R-P03), ``It would be important to talk to a board-certified medical oncologist to determine and discuss all the pros and cons of various different regimens, some of which are more toxic, and some of which are much less toxic and well tolerated.” (R-P04), ``So that would be something that would absolutely need to be discussed with a primary care doctor and likely with a cardiologist to ensure that your sister's risk of having a heart attack in the future is well controlled and treated.” (R-P01), ``With all these imaging data we actually have a very regular multidisciplinary team of doctors including myself, as well as thoracic surgeons, radiation oncologists, pathologists, pulmonologists, and oncologists all gathered together on a weekly basis to discuss the cases because there's not only diagnostic uncertainty but there are also treatment options that vary in their profile and some things are more appropriate than others.”(R-P1)}

\paragraph{\textbf{Theme 3}: Locating Issues mentioned in the Radiology Report to the CT Scan} included discussions of how the participant could locate a specific term from the radiology report on the CT scan. This information was also brought up by the radiologist himself in answering a question where pointing to the image might be helpful. The radiologist annotated the on-screen scan to demarcate parts of the scan. For example, \textit{ ``If you look at the scans in image number 1 and 2, you can see that there are kind of dark bubbly spots and I can use the zoom technology to annotate for you. I'll just show an example area here. Other example areas like here are kind of dark bubbly spots that are present in the lungs and that represents areas of just ballooned out Airways as a result of destruction.” (R-P02), ``So how do I find a nodule here?” (P04), ``If I can just sort of highlight that area here. That and that. Do you see these round sort of white dots, white sort of areas? These are the areas that I'm concerned might be because of lung cancer.” (R-P02), ``How would I differentiate between the scan of a healthy lung versus the one which is in stage one, stage 2, or stage 3? So, looking at it, what should I look for?” (P04)}

\paragraph{\textbf{Theme 4}: Selecting a Treatment Plan} included a range of discussions comparing and contrasting the treatment plans available in terms of their success rate, degree of discomfort to the patient, and the costs involved. Examples are as follows –  \textit{``As long as she's in relatively and reasonably good health to walk around, typically surgery is an option and the exact evaluation of which I defer to my thoracic surgery colleagues to do the evaluation.” (R-P03), ``If the cancer has already spread throughout the body, then chemotherapy may be the only option available to shrink the cancer and seek care.” (R-P04), ``Between chemotherapy or radiation, which has the least side effects, because when I see some patients going through the treatment, the way they suffer, sometimes even they themselves feel that they would rather die than go through the treatment.” (P04), ``What treatment option would also say is like more expensive than others and vice versa?”(P03)}

\paragraph{\textbf{Theme 5}: Understanding Medical Terminology from Radiology Report} dealt understanding the content of the radiology report. For example,  \textit{``Got it, I know you mentioned emphysema. What exactly does that mean? Is that just like lung, you know, like problems with, like the lung or like lung being cancerous, is that but that stands for?” (P03), `` For some terminology that I wasn't entirely clear on, like the pleura and then mediastinum, nothing special was mentioned, but can you just briefly mention what those mean and what you were looking for there?” (P05), ``Should I be also worried about these Hepatic hyperdensities in the visible abdomen area?” (P1), ``The one main finding is the lung RADS category 4B that's mentioned in the impression section and what that is referring to is based on the two lung nodules, and these are sort of tiny, sometimes very tiny lesions that are found in the lungs, many of which can be a consequence of having infection in the past or just air pollution or other things.”(R-P02)}

\paragraph{\textbf{Theme 6}: Diagnostic Follow-up Plan}  included content related to the next diagnostic steps to be taken following the initial CT scan – discussions on the different kinds of follow-up options available and what would be the most appropriate in this given situation. E.g, –  \textit{``Tissue sampling means biopsy – so we insert a needle either through the skin or using a bronchoscope and then take a surgical sample of these lesions and then take a look at these them under the microscope to tell for sure.” (R-P03), ``In the case of PET CT, it's a metabolic imaging, so it highlights areas of high metabolic activity, which often corresponds to areas of cancer.” (R-P03), ``Would these scans happen pretty sequentially in just like a couple of weeks? Or would we have months between these scans or how long I guess would this last?” (P03), ``So it seems like there is definitely a suspicion of lung cancer. What would you recommend to be the next steps? A chest CT with contrast has been recommended from the results. But how is that different from this and what would that reveal? (P03)”}

\paragraph{\textbf{Theme 7}: Risk Factors and Lifestyle Changes}  included content related to different kinds of risk factors involved in lung cancer, and discussions on whether certain lifestyle factors might help curb the progression of the disease. For example,  \textit{``Exercise has been shown to be helpful in numerous studies to reduce the risk of cardiovascular disease and would certainly be recommended assuming it's not just overdone and there are no conditions that cause issues with exercising.” (R-P1), `` Most commonly in the United States, it's because of smoking, but it can also be due to air pollution or some other conditions that may result in the destruction of that part of the lungs.” (R-P02), ``If smoking is continuing, making sure to stop smoking is important. Also if there's some sort of pollutant or other thing, if there's any way to really just keep your lungs healthy from here on, it will be important to make sure that it doesn't progress further and cause more shortness of breath.” (R-P1), ``Are there some foods that kind of help with these things? Or in general, is weight loss better?” (P1)}

\paragraph{\textbf{Theme 8}: Disease Progression and Prognosis} included content about the way the disease was progressing, and what the patient could expect in the future. For example,  \textit{``I was mostly concerned about the health risks going forward and whether there was any way we can kind of stop the cancer from happening in some way.” (P1), ``If you were to follow this module, say in three months, and re-scan if this actually was truly a cancer, then it would grow in a pattern and growth rate that's typical of a lung cancer. Whereas if this represents something else and something else could be, say, the focus of infection or inflammation, these tend to either go away or increase or change in a pattern and growth rate that's not consistent with cancer.” (R-P03), ``Assuming that this nodule turns out to be cancer, it would be stage one A or one B, which is a really good stage to be in because I don't see any nodal metastasis.” (R-P04), ``Can they shrink, or can you make them disappear, such that in the follow-up scan, you will have a cleaner scan? Is this possible?” (P05)}

\paragraph{\textbf{Theme 9}: Diagnostic Confidence and CT Scan Limitations} dealt with discussions on how confident the radiologist was about the findings and the general limitations of the presented scans. Examples include,  \textit{``Having said that, there is some diagnostic uncertainty with these that can be more definitively addressed by doing a follow-up CT scan or doing some additional imaging modalities or doing a biopsy.” (R-P1), ``I will add one caveat, we only scan the chest. We didn't scan the brain. We didn't scan the abdomen. So sometimes the cancers surprise us and say, without having any adjacent lymph, node metastasis, we see the first metastasis in the brain.”(R-P04), ``I personally would probably classify the prob of these being cancer somewhere between 20 to 50 percent.”(R-P1), ``I just want to mention again that there is some diagnostic uncertainty at this point. This could be cancer. This could also be an infectious process or something unrelated.”(R-P1)}

\paragraph{\textbf{Theme 10}: Alleviating Patient Anxiety} included content that specifically dealt with managing participants’ emotional responses to the diagnoses. Examples of quotations from this theme are as follows –  \textit{``Now, I don't want to have you get so worried right away immediately, because even though they raise suspicion and possibility of lung cancer. It is true many of them also don't turn out to be cancer.” (R-P04), ``It's very, very common for people, patients at age 80s to have coronary artery disease. So I wouldn't necessarily be so shocked about it and so deeply concerned about it.”(R-P03), ``And so in general, if I were to discuss this with my dad and he asked me how concerned should I be in general at this stage you know. What should I tell him, based on these findings?”(P05), ``First of all, concern, and anxiety is never really a good thing in the medical diagnostic and treatment process. But that's very understandable with just receiving the impression that there is a suspicious lesion. Now, we don't know for sure that this is cancer. But this may turn out to be lung cancer, in which case there are some further steps. But what I usually tell patients is that for now work with us in the medical system and we will go through excellent, standard steps of care.”(R-P05)}

\subsection{Frequency of Themes across Radiologist-Patient Interactions}
\begin{figure}[t]
    \centering
    \includegraphics[width=0.9\linewidth]{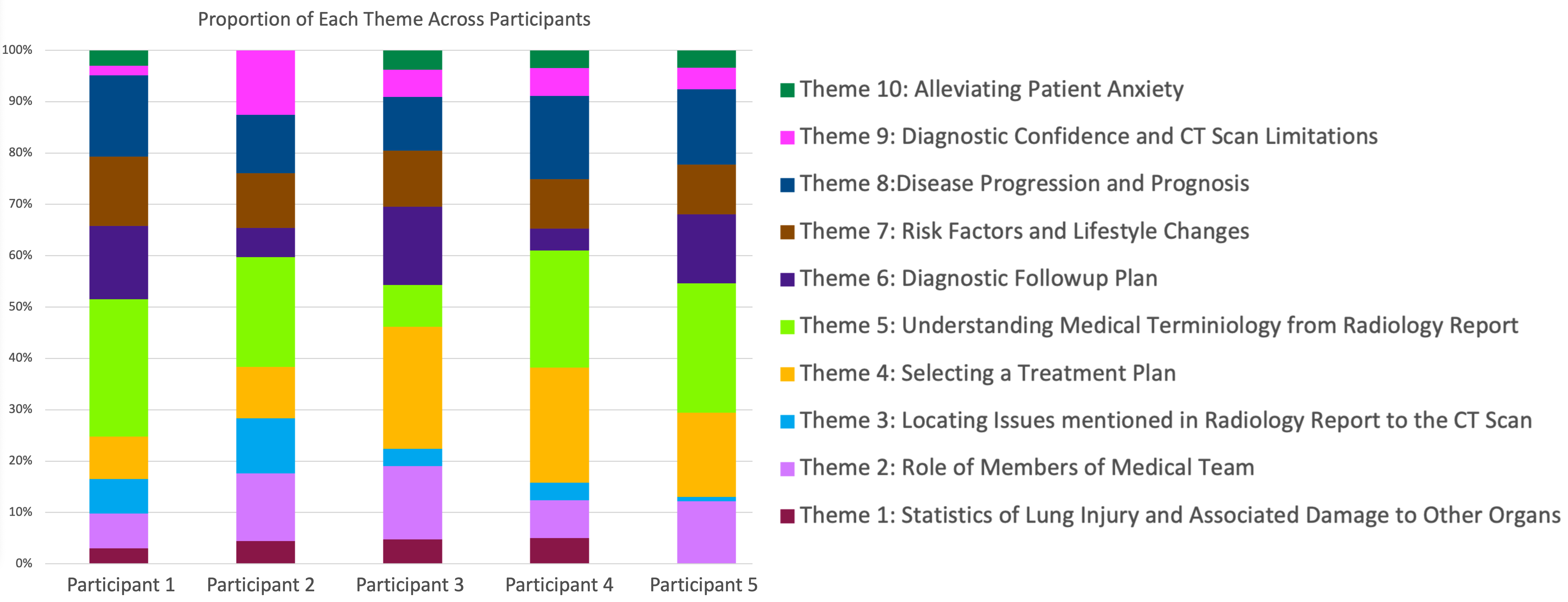}
    \caption{Stacked bar chart describing the proportion of each theme in each participant’s interaction with the radiologist, with the portion of the conversation occupied by each theme on the y-axis, and participants along the x-axis.}
    \label{fig:3}
\end{figure}

Figure \ref{fig:3} depicts the proportion of various themes in each participant’s interaction with the radiologist. Although there are variations in how much each participant’s conversation incorporates each theme, there appear to be several consistencies across participants. The following are a few noteworthy observations:
\setlist{nolistsep}
\begin{itemize}[noitemsep,left=0pt]
\item Largely, all themes seem to be represented across the radiologist interactions of all 5 participants. The exceptions are Participant 2, who is missing the \textit{Alleviating Anxiety} theme, and Participant 5, who is missing the \textit{Statistics of Lung Injury and Associated Damage to Other Organs theme}. This is a good indication that we have identified a comprehensive, well-rounded set of themes that will likely suffice to describe interactions with any new participants.
\item The \textit{Understanding Medical Terminology from Radiology Report} and \textit{Selecting a Treatment Plan} themes constitute the largest portion of the interactions. The emphasis on understanding terminology suggests that individuals are the most keen to move past the complex medical terminology and want to understand what the report means in accessible language. The emphasis on selecting treatment plans likely suggests that participants in the task were able to incorporate feelings of empathy for their imagined relative’s condition, and are extremely interested in understanding treatment options and performing a cost-benefit analysis.
\item \textit{Alleviating Anxiety} and \textit{Statistics of Lung Injury and Associated Damage to Other Organs} themes constitute the smallest portions of the interactions. This is interesting because participants seem minimally interested in statistical facts about the likelihood of different aspects of the disease, and also in emotionally soothing statements by the radiologist that only serve the purpose of alleviating anxiety.
\item \textit{Disease Progression and Prognosis},\textit{ Risk Factors and Lifestyle Changes, Diagnostic Followup Plan}  and \textit{Role of Members of the Medical Team} all comprise fairly similar proportions of the interactions across participants. There seems to be a significant level of curiosity regarding what caused the disease and what the next steps are. There also seems to be considerable discussion on the roles of different members of the medical team, since participants often ask questions that are beyond the scope of the radiologist’s role.
\item \textit{Locating Issues Mentioned in the Radiology Report in the CT Scan} comprises a surprisingly small proportion of all interactions. While participants are keen to understand what the report means, and what treatment options their relative will have, they seem to have little interest in understanding how the disease presents in the scans.
\end{itemize}

\section{Generative AI Evaluation}
Successful adoption of generative AI systems in patient-facing applications requires that they must respond reliably, factually, and meaningfully to questions asked by patients. In this section, we evaluate two state-of-art, commercially-deployed generative AI (GenAI) systems : \textsc{Claude 3.5 Sonnet}~\cite{claude2024}  and \textsc{ChatGPT 4o}~\cite{gpt2024}. Both of these systems are built upon language models that are top-scores in standardized multi-task understanding benchmarks such as MMLU-Pro~\cite{wang2024mmlu}. MMLU-Pro is a question-answering dataset containing $12032$ questions and multiple choice answers from health, psychology, and biology domains in addition to other fields. Our goal is to explore if these highly successful models can address questions humans ask naturally as they perform a sensemaking task.    

\subsection{Study}
\subsubsection{Conversational MultiModal Dataset} We, first, built a small multimodal dataset of question-answer pairs from the formative interview study and corresponding thematic analysis \ref{sec:study}. Our dataset represents $5$ themes; 
\begin{enumerate}
    \item Theme 3: locating issues mentioned in the radiology report to the CT scan,
    \item Theme 4: selecting a treatment plan,  
    \item Theme 5: understanding medical terminology from radiology reports,
    \item Theme 6: diagnostic follow-up plan, and
    \item Theme 7: risk factors and lifestyle changes. 
\end{enumerate}
We chose these $5$ themes because they allowed for the clearest formulation of meaningful question-answer pairs. The remaining themes were largely observed in parts of the radiologist's explanations to answers to questions from the selected themes. For each theme, we selected $5$ questions that were representative of the diversity in the questions study participants asked. We paired each question with the long-form answer that the radiologist provided during the interview study.  

We structured the evaluation dataset as a set of $25$ visual question-answering problems. Each problem consisted of (1) two images from radiology scans (in Figure \ref{fig:case-report}), (2) corresponding radiology report capturing expert impressions (in Figure \ref{fig:case-report}), and (3) a question asked by the patient. 
This setup requires the GenAI system under study to perform visual reasoning on the CT scan images and integrate that evidence with analytical understanding of the report to generate long-form answers.

\subsubsection{Protocol} We accessed both \textsc{Claude-3.5-Sonnet} and \textsc{ChatGPT-4o} through their web interfaces. Generative AI systems are malleable and their behavior can be modulated through system instructions and various prompting techniques. Several of these prompting techniques (few-shot prompting, chain-of-thought reasoning) require the human to have expertise in interacting with generative AI systems and an understanding of their behavior. Patients are unlikely to have this expertise. Consequently, we studied two configuration that are most natural in our setting. In the first configuration (/config1), the system is provided with all background material (two scans and the report) and asked a question. This configuration reflects how a patient might use the publicly available GenAI systems. In the second configuration (/config2), we added a system prompt to guide AI's behavior. The system prompt provides the context in which the question is asked and includes who is asking the question (\emph{who}: a patient's caregiver), what the questioner is looking for (\emph{why}: they want to understand the scans and reports), and how the response should be generated (\emph{how}: simple, non-clinical language, and a succinct answer). The configuration instructs GenAI systems that the questioner doesn't have a background in clinical practice, biasing them to use colloquial language and descriptions. This configuration reflects the guidance a GenAI system might receive from a healthcare system administrator as the system is deployed.  

Both GenAI systems were tested under the two configurations on all $25$ evaluation questions and their answers were recorded. After the system answered a question, the conversation history was reset. This was done to ensure independence in response generation. GenAI systems are autoregressive and generate answers by sampling from a token distribution based on which tokens were generated previously in the conversation. Resetting the history ensures that incorrect information in earlier response do not bias response generation for the question under study. 

We complied the questions, expert answers, and responses generated by both configurations of the two GenAI systems understudy in an evaluation dataset\footnote{Raw dataset is available at \url{https://docs.google.com/spreadsheets/d/e/2PACX-1vTZB2sw1JadKKc1TXF8dzF-WYL7cOFta6E61_IIdyVvR1ncDzm0nceFpA3JFPwZ5acWVjDM8hpYB9Ki/pubhtml}}. 

\subsection{Analysis}
\subsubsection{Empirical Findings}
Table \ref{tab:word-count} shows the average length of expert responses as compared to those produced by the GenAI systems under both experimental conditions. Natively, both \textsc{Claude-3.5-sonnet} and \textsc{ChatGPT-4o} tend to be verbose and produce significantly longer responses. However, when they are configured with the system prompt encouraging them to produce answers that are succinct, the response length reduces reliably for both. 

\begin{table}[]
\caption{A table noting the average length of responses from a radiologist and variations of GenAI models: \textsc{Claude} and \textsc{ChatGPT}}
\begin{tabular}{@{}lc@{}}
\toprule
Response type   & Avg. words (std. dev.) \\ \midrule
Radiologist     & 146.2 (82.68)          \\
Claude/config1  & 281.0 (35.85)          \\
Claude/config2  & 167.88 (79.22)         \\
ChatGPT/config1 & 178.36 (53.42)         \\
ChatGPT/config2 & 134.12 (45.77)         \\ \bottomrule
\end{tabular}
\label{tab:word-count}
\end{table}

Next, we evaluate if the GenAI system responses are similar to how a radiologist responds. Evaluating similarity of free-form text is a challenge in itself. Classical measures such as ROUGE score \cite{lin2004rouge} rely on overlap in the words used in two pieces of text. However, difference in stylistic choices may result in lower scores even when the two texts have similar meaning. 

To compare meaning-level similarity between expert and GenAI responses, we use the universal sentence encoder (USE) that encodes text into high dimensional vectors~\cite{cer2018universal}. The model is trained and optimized for greater-than-word length text, such as sentences, phrases or short paragraphs making it suitable for our purposes. It is trained on a variety of data sources and a variety of tasks with the aim of dynamically accommodating a wide variety of natural language understanding tasks. The input is variable length English text and the output is a $512$ dimensional vector. Once the responses are converted into high-dimensional vectors, their cosine similarity ranges from $0$ to $1$ indicating how close they are in the semantic space, $1$ when the responses are the same. In other words, the cosine similarity indicates the degree to which a GenAI system's response is aligned with an expert. 

\begin{figure}
    \centering
    \includegraphics[width=1\linewidth]{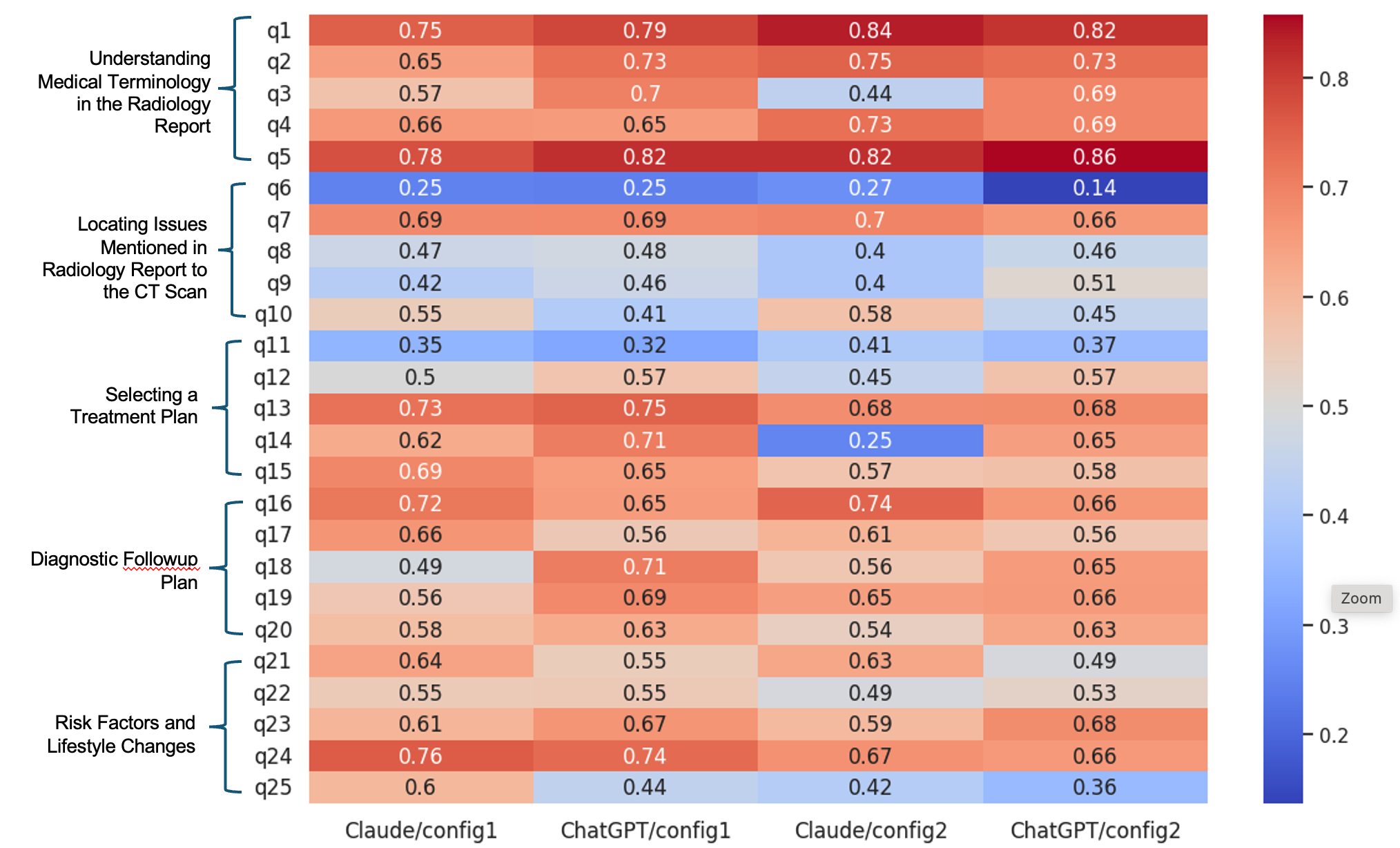}
    \caption{A heatmap plotting the semantic similarity of responses generated by Claude and ChatGPT under two experimental configurations to an expert's response.}
    \label{fig:similarity}
\end{figure}

Figure \ref{fig:similarity} shows how similar the responses generated by the experimental configurations of GenAI systems are to those of a human expert. The scale goes from dark blue to dark red (dissimilar to similar). We make the following observations. 
\begin{enumerate}
    \item There is a large variability in measured similarity. This variability indicates that generative AI systems are not in full alignment with the expert radiologist. In a majority of the cases, their responses are different from that of a radiologist scoring less than $0.75$ on cosine similarity. 
    \item Response similarity varies between the themes. Highest similarity are observed in \textbf{Theme 5: Understanding Medical Terminology in the Radiology Report}. This is not surprising given that clinical textbooks are included in the datasets LLMs are trained over. Lowest similarity responses are observed in \textbf{Theme 3: Locating Issues Mentioned in Radiology Report to the CT scan}. We investigate the reason for this observation in our qualitative analysis (\ref{sec:section_themes}).
    \item Configuration 2 that encourages the system to produce short text accessible to non-clinicians appears to reduce the similarity measure. Mostly, the impact is small. However in some cases (e.g, q3, q14 Claude; q21, q22 ChatGPT) it is especially pronounced. Reduction in similarity indicates that GenAI system do not know how to focus the generative process on relevant information. They can remove information that the radiologist considers important to communicate. 
\end{enumerate}

\subsubsection{Qualitative Investigation} 
Similarity scores reported earlier indicate that there are differences between how a radiologist responds to patient queries and how a GenAI system approaches answer generation. However, the scores themselves don't explain why or how these answers are different. In this section, we qualitatively analyze the difference between the responses provided by the radiologist from those generated by the GenAI systems.

Drawing inspiration from the Griean maxims \cite{grice1975logic} of effective conversation, we identify the following dimensions of evaluation. 

\begin{figure}
    \centering
    \includegraphics[width=1\linewidth]{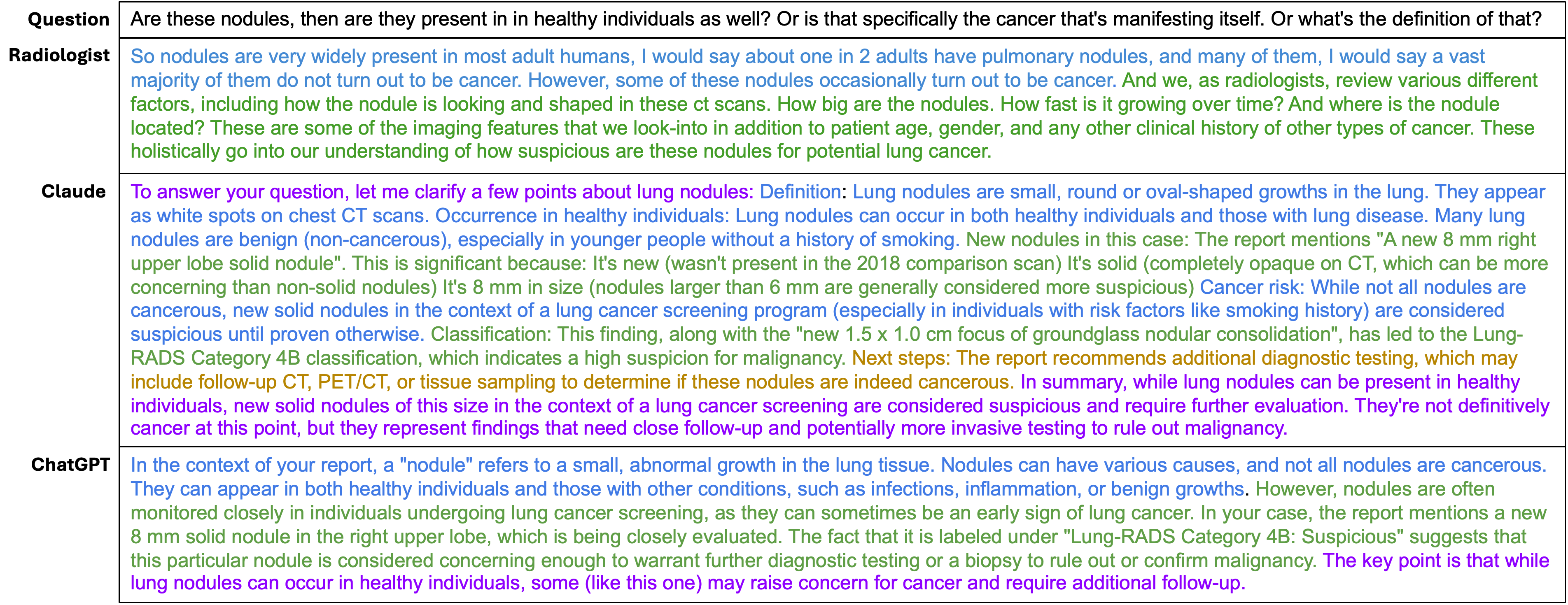}
    \caption{Theme 5 question (id=q1) and responses by the expert and GenAI systems. Blue denotes sentences that contain a direct response to the question; green, relevant elaborations; orange, irrelevant elaboration; and purple, superfluous text.}
    \label{fig:response}
\end{figure}

\begin{enumerate}
    \item \emph{Correctness} follows the Gricean maxim of quality, where one tries to be truthful, and does not give information that is false or that is not supported by evidence. We engaged an expert evaluator (a radiologist and a co-author on this paper) to judge if the responses produced by the GenAI systems were correct. We adopted a particularly strict criterion for correctness - whenever the expert identified even a single non-factual sentence in the response, it was marked to be incorrect. Such strict evaluation is necessary for critical applications of generative AI models that have been known to hallucinate information which may be dangerous in our usecase.
    \item \emph{Relevance and informativeness} follow the Gricean maxim of relation (where one tries to be relevant, and says things that are pertinent to the discussion) and quantity (where one tries to be as informative as one possibly can, and gives as much information as is needed, and no more). We characterized each sentence in the answers as: \emph{direct response}, sentences that contain a clear answer to the question asked; \emph{relevant elaboration}, sentences that elaborate the answer by providing additional, supporting information; \emph{irrelevant elaboration}, sentences that contain information irrelevant to the question asked; and finally, \emph{superfluous}, sentences with little information about the question. Figure \ref{fig:response} shows an example of such characterization where text in blue indicates direct response, green - relevant elaboration, orange - irrelevant elaboration, and purple - superfluous text. The characterization was done by authors who are not medical experts and appropriate judges to evaluate if the answers are relevant and informative. 
\end{enumerate}

\paragraph{Finding 1} On the correctness dimension, we found that out of $25$ questions, \textsc{Claude/config1} was judged to be incorrect on $10$ ($40\%$ error rate) while \textsc{ChatGPT/config1} on $5$ ($20\%$ error rate) leading the expert to reject their answers. In two instances the expert accepted the responses while noting that the response didn't answer the question but was correct. Errors manifested due to a variety of causes including: inablility to relate causal factors to diseases, incorrectly identifying risks and consequences, inability to distinguish generic textbook information from specifics of the case, and factual incorrectness due to hallucinations. The expert noted that the GenAI systems ramble on and often, include information that is not directly relevant to the question.

\paragraph{Finding 2} Our analysis of the relevance \& informativeness dimension revealed several insights about how doctors address their patient's questions. The radiologist's responses consisted of sentences that directly answer the patient question followed by a relevant elaboration. An example is shown in Figure \ref{fig:response} where after explaining that the majority of nodules do not turn out to be cancer, the doctors describes the criteria that are used to evaluate nodules. Other elaborations included demonstrations, impact of the disease the patient can expect to see, what further evidence may manifest, how a diagnosis may be made, and how a treatment plan may be developed. This finding hints at the mixed-initiative nature of the patient-provider dialog where the role of the provider is not just to answer questions but also educate the patient so that they feel empowered to take informed decisions about their health. We didn't observe any other information categories in the radiologist's response. 

Both \textsc{Claude-3.5-Sonnet} and \textsc{ChatGPT-4o} produced long answers with elaborations. An example is shown in Figure \ref{fig:response} where the relevant elaborations are shown in green and the irrelevant ones in orange. Explanation of how nodule's size or appearance influences the diagnosis of cancer is relevant for the patient. However, next steps are not pertinent to the question asked. Additionally, both generated some superfluous information shown in purple preface or summarize its actual response but doesn't add any useful information to the response. Production of irrelevant elaborations and superfluous information directly violates of desiderata of relevance \& informativeness and the Gricean maxims of relation and quantity.

\begin{figure}[t]
    \centering
    \includegraphics[width=1\linewidth]{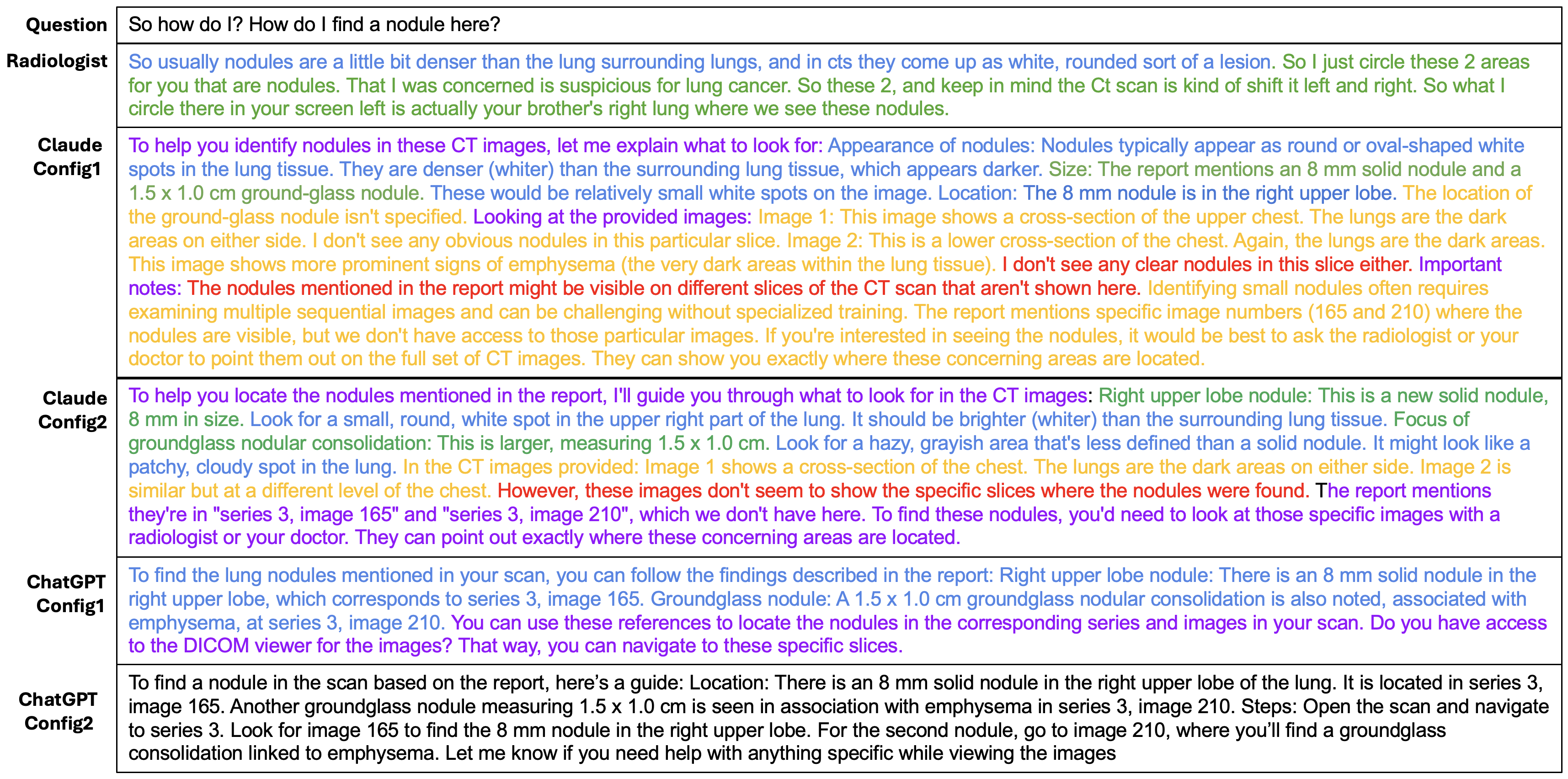}
    \caption{Theme 3 question (id=q7) and corresponding answers provided by the radiologist and GenAI systems}
    \label{fig:response2}
\end{figure}

\paragraph{Finding 3} Both GenAI systems struggled with responding to questions in theme 3: \emph{locating issues mentioned in the radiology report to the CT scan}, explaining the low similarity scores in Figure \ref{fig:similarity}. An illustrative example is shown in Figure \ref{fig:response2}. To answer the question, the radiologist responded by first describing how to find relevant areas - nodules - in the scan and then, a demonstrative elaboration highlighting those in the scan to ground the caregiver's understanding. 

The demonstrative elaboration was copiously missing in both \textsc{Claude}/config1 and \textsc{ChatGPT}/config1 responses. They both generated generic guidance on how to locate the nodule in the scan and regurgitated language about nodules from the scan. It is to be expected. GenAI systems are not designed to guide the questioner's attention to visual evidence contributed to answer generation. Further, \textsc{Claude}/config1 and \textsc{Claude}/config2 incorrectly believe that the nodule cannot be localized in the scans. Our observations call into critique the assumed 'multi-modality' of GenAI system when these systems only consume visual information to guide response generation but cannot use the visual medium to communicate information. 

\textsc{ChatGPT}/config2 generated an answer that is correct however, is entirely useless to the patient. It present banal/obvious information about how to localize the nodule. This finding indicates that we should revise our relevance and informativeness criteria to include banality. 

\begin{figure}[t]
    \centering
    \includegraphics[width=1\linewidth]{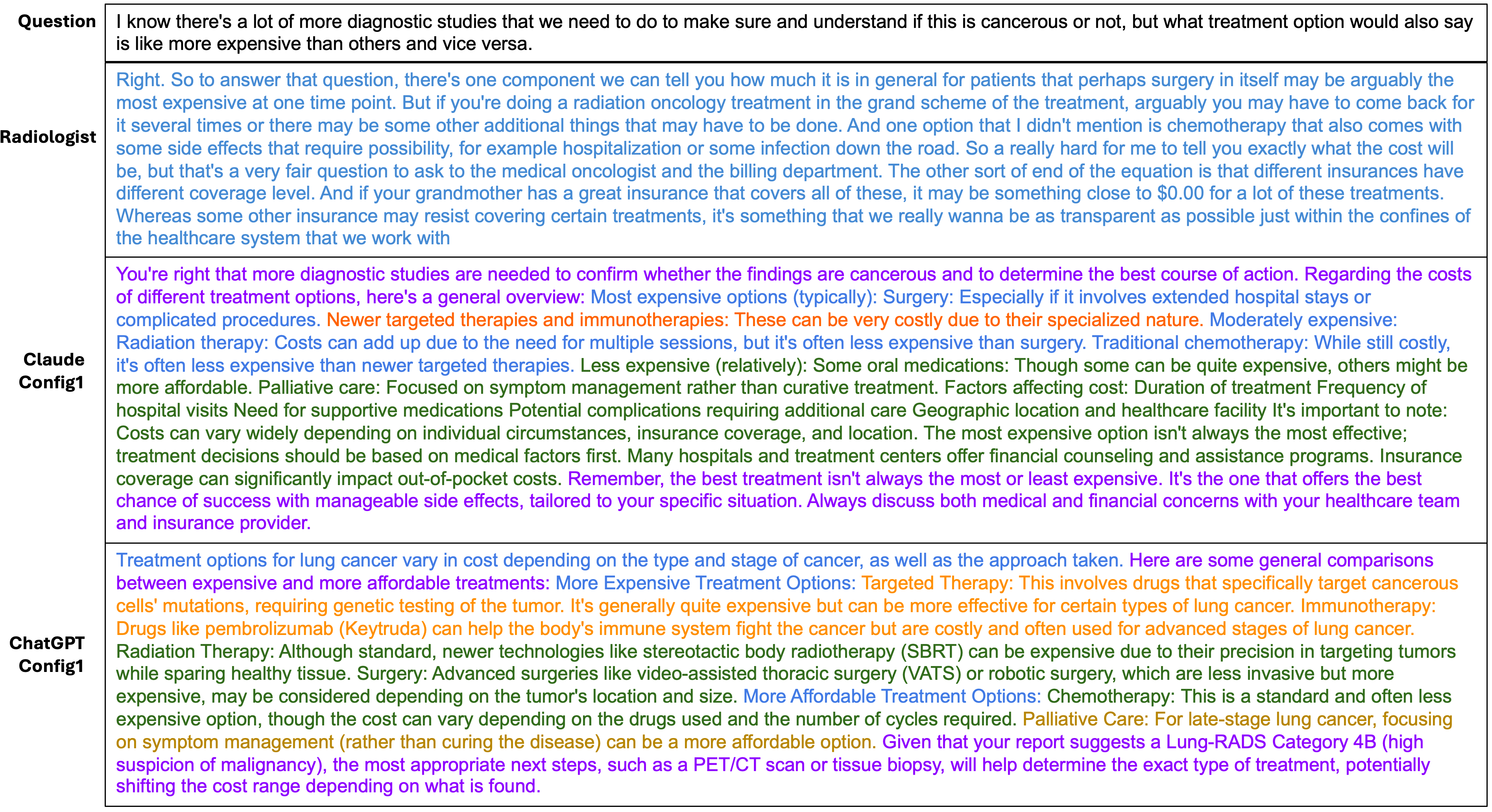}
    \caption{Theme 6 question (id=q15) and corresponding answers provided by the radiologist and GenAI systems}
    \label{fig:response3}
\end{figure}

\textsc{Finding 4} In several responses, the radiologist went beyond the exact question being asked and answered a higher level concern. This approach was particularly visible in \emph{Theme 6: diagnostic follow-up plan}. An example is shown in Figure \ref{fig:response3} where the patient enquires about expected expenses for various treatment options. The radiologist's response not only compares different treatment options in terms of how expensive they are, it also provides them with other tradeoffs that the patient didn't explicitly inquire about. Examples include observations such as radiation oncology requires the patient to come back several times and chemotherapy has side-effects that might require hospitalization. In contrast, both \textsc{Claude-3.5-Sonnet} and \textsc{ChatGPT-4o} focused on costs associated with treatment options, often elaborating to include less common or new therapies. In general we found the radiologist's responses were geared towards helping the patients make decisions whereas GenAI systems' responses were geared towards extensive enumeration of possibilities and definitions.    

\textbf{Finding 5} A meta-finding of our research is that the USE-based similarity metric isn't sufficiently selective for our tasks. Consider the example in Figure \ref{fig:response2} where it is easy to see that the radiologist's responses in grounded in the CT scans and discusses localization of nodules. GenAI systems on the other hand produce responses containing different information. However, their similarity scores (Figure \ref{fig:similarity}, line q7) are almost $0.7$ indicating a good degree match. This observation indicates that USE supports a topic-level similarity but cannot compare the content as a human would, despite being one of the standard measurement techniques in AI.  
\section{Discussion, Conclusions, and Future Work}
In this paper, we study if current state-of-art in generative AI systems can support patients (and caregivers) as they attempt to understand their scans and reports. To structure our evaluation, we conducted a need-finding study of caregiver-radiologist interactions. We identified $10$ themes of diverse information types that such interactions are typically comprised of. We used $4$ themes to create an evaluation question-answer dataset and evaluated two state-of-art models: \textsc{Claude-3.5-Sonnet} and \textsc{ChaptGPT-4o} both quantitatively, by measuring response similarity with a human expert and qualitatively, by judging the responses on Gricean maxims of effective communication. 

Our findings indicate that despite significant optimism about generative models in medical contexts, the current state-of-art doesn't sufficiently address patients' information needs. Surprisingly, we found that these models have high error rates ($20\%$ for ChatGPT-4o, $40\%$ in Claude-3.5-Sonnet) in our evaluation set built on question asked in real interactions. This finding is concerning - patients are not experts of medicine and consequently, cannot identify misinformation in AI-generated responses. Any system that is deployed must guaranteed to provide factual information. The models were also deficient as conversational information systems and frequently produced irrelevant and superfluous elaborations violating Gricean maxims. This finding suggests that it is unlikely that patients will be able to use this system to access useful information. 

Our findings also highlight a critical gap in the evaluation paradigm prominent in AI \& ML and the challenges of accountable deployment of generative AI systems. In  machine learning literature \cite{bubeck2023sparks}, these models are shown to have high accuracy on curated, task-agnostic datasets and claimed to have human-level performance. However, to be at par with human experts, the models not only have to answer questions but also operationalize this knowledge in practical, real-world tasks such as helping patients understand their illness and determine a treatment plan - a task doctors and healthcare worker routinely perform. Our work paves the way for task-specific evaluation protocols that can spur the generative AI movement in a beneficial direction. 

There are several avenues for future work. The study described here is small (only $5$ participants) and limited to a single case of radiological imaging. Consequently, the themes identified here do not represent an exhaustive set of informational needs patients have. In doing this arguably limited study, we formulated a concrete approach of evaluating GenAI systems. Building upon our learnings, we will expand the study to include a larger, heterogeneous population as well as cover a greater variation of medical contexts. A larger study will enable us to create a comprehensive evaluation dataset for GenAI systems, ensuring a rigorous task-centric evaluation of GenAI systems. 

The second avenue is studying a wider range of GenAI system configurations. Generative AI systems are considerably malleable and their behavior can be modulated through various prompting strategies, supervised fine tuning, and reinforcement learning etc. This paper focuses on sketching out an evaluation paradigm by studying two simple configurations. In future, we will build and study methods that align GenAI system response to experts.    

Third, prominent metrics such like BLEU~\cite{papineni2002bleu}, ROGUE~\cite{lin2004rouge}, BERT Score~\cite{zhang2020bertscore}, or USE~\cite{cer2018universal} while commonly used to evaluate LLM responses, fall short of measuring if the response produced are human-like or if they adequately address the humans' information needs. To address this shortcoming in AI \& ML evaluation paradigm, we will develop a quantitative measurement scale based on our the qualitative evaluation paradigm proposed here. Finally, we are invested in developing intelligent assistive systems for the healthcare. Our future work will explore how the shortcoming highlighted in this paper can be alleviated with modern and classical AI methods to develop useful conversational agents. 

Despite its limitations, our makes several important contributions towards sustainable and productive deployment of generative AI systems. First, the performance of generative AI systems is typically measured on domain-agnostic, task-agnostic question-answering datasets. While these methods are useful in comparing algorithms against each other, they provide little insight about how useful the answers are for humans. Our research situates question answering in a specific context of people seeking information to enhance understanding in healthcare settings. This enables us measure the degree to which GenAI systems are useful. Second, GenAI systems are typically evaluated on datasets curated by and for clinicians. These datasets contain questions and answers that are most interesting to clinicians while performance measurement includes criterion such as diagnostic accuracy. Such an approach introduces an implicit bias in the design and analysis of these systems - they are designed to replace clinicians by mimicking the output of their problem-solving process, following the \emph{AI-is-automation} philosophy. However, to design a good information support tool, paying attention to who the seeker and the consumer of that information is critical. Patients and caregivers have vastly different needs than clinicians who are experts of terminology, diagnostics, and treatment. Our research takes the \emph{AI-is-augmentation} view and centers patients and caregivers, adopting a much needed lens missing from the medical generative AI research landscape. Finally, this paper lays the groundwork for a principled, within-context evaluation approach for generative AI systems. We study the space of information needs of a patient/caregiver to understand what types of questions are asked by the patient/caregiver. We introduce a novel qualitative answer evaluation paradigm that measures both correctness and relevance of answers generated by AI systems. Our research can be considered a precursor to a rigorous, quantitative approach for measuring suitability of a generative AI system for addressing humans' informational needs.

\bibliography{bibliography}

\begin{thebibliography}{55}
\expandafter\ifx\csname natexlab\endcsname\relax\def\natexlab#1{#1}\fi
\providecommand{\url}[1]{\texttt{#1}}
\providecommand{\href}[2]{#2}
\providecommand{\path}[1]{#1}
\providecommand{\DOIprefix}{doi:}
\providecommand{\ArXivprefix}{arXiv:}
\providecommand{\URLprefix}{URL: }
\providecommand{\Pubmedprefix}{pmid:}
\providecommand{\doi}[1]{\href{http://dx.doi.org/#1}{\path{#1}}}
\providecommand{\Pubmed}[1]{\href{pmid:#1}{\path{#1}}}
\providecommand{\bibinfo}[2]{#2}
\ifx\xfnm\relax \def\xfnm[#1]{\unskip,\space#1}\fi
\bibitem[{Dwivedi et~al.(2023)Dwivedi, Kshetri, Hughes, Slade, Jeyaraj, Kar,
  Baabdullah, Koohang, Raghavan, Ahuja et~al.}]{dwivedi2023so}
\bibinfo{author}{Y.~K. Dwivedi}, \bibinfo{author}{N.~Kshetri},
  \bibinfo{author}{L.~Hughes}, \bibinfo{author}{E.~L. Slade},
  \bibinfo{author}{A.~Jeyaraj}, \bibinfo{author}{A.~K. Kar},
  \bibinfo{author}{A.~M. Baabdullah}, \bibinfo{author}{A.~Koohang},
  \bibinfo{author}{V.~Raghavan}, \bibinfo{author}{M.~Ahuja}, et~al.,
\newblock \bibinfo{title}{{“So What if ChatGPT Wrote it?” Multidisciplinary
  Perspectives on Opportunities, Challenges and Implications of Generative
  Conversational AI for Research, Practice and Policy}},
\newblock \bibinfo{journal}{International Journal of Information Management}
  \bibinfo{volume}{71} (\bibinfo{year}{2023}) \bibinfo{pages}{102642}.
\bibitem[{Jin et~al.(2021)Jin, Pan, Oufattole, Weng, Fang, and
  Szolovits}]{jin2021disease}
\bibinfo{author}{D.~Jin}, \bibinfo{author}{E.~Pan},
  \bibinfo{author}{N.~Oufattole}, \bibinfo{author}{W.-H. Weng},
  \bibinfo{author}{H.~Fang}, \bibinfo{author}{P.~Szolovits},
\newblock \bibinfo{title}{What disease does this patient have? {A}
  {L}arge-scale {O}pen {D}omain {Q}uestion {A}nswering {D}ataset from {M}edical
  {E}xams},
\newblock \bibinfo{journal}{Applied Sciences} \bibinfo{volume}{11}
  (\bibinfo{year}{2021}) \bibinfo{pages}{6421}.
\bibitem[{Cascella et~al.(2023)Cascella, Montomoli, Bellini, and
  Bignami}]{cascella2023evaluating}
\bibinfo{author}{M.~Cascella}, \bibinfo{author}{J.~Montomoli},
  \bibinfo{author}{V.~Bellini}, \bibinfo{author}{E.~Bignami},
\newblock \bibinfo{title}{Evaluating the feasibility of chatgpt in healthcare:
  an analysis of multiple clinical and research scenarios},
\newblock \bibinfo{journal}{Journal of medical systems} \bibinfo{volume}{47}
  (\bibinfo{year}{2023}) \bibinfo{pages}{33}.
\bibitem[{Clusmann et~al.(2023)Clusmann, Kolbinger, Muti, Carrero, Eckardt,
  Laleh, L{\"o}ffler, Schwarzkopf, Unger, Veldhuizen
  et~al.}]{clusmann2023future}
\bibinfo{author}{J.~Clusmann}, \bibinfo{author}{F.~R. Kolbinger},
  \bibinfo{author}{H.~S. Muti}, \bibinfo{author}{Z.~I. Carrero},
  \bibinfo{author}{J.-N. Eckardt}, \bibinfo{author}{N.~G. Laleh},
  \bibinfo{author}{C.~M.~L. L{\"o}ffler}, \bibinfo{author}{S.-C. Schwarzkopf},
  \bibinfo{author}{M.~Unger}, \bibinfo{author}{G.~P. Veldhuizen}, et~al.,
\newblock \bibinfo{title}{The future landscape of large language models in
  medicine},
\newblock \bibinfo{journal}{Communications medicine} \bibinfo{volume}{3}
  (\bibinfo{year}{2023}) \bibinfo{pages}{141}.
\bibitem[{Adams(2010)}]{adams2010improving}
\bibinfo{author}{R.~J. Adams},
\newblock \bibinfo{title}{Improving {H}ealth {O}utcomes with {B}etter {P}atient
  {U}nderstanding and {E}ducation},
\newblock \bibinfo{journal}{Risk management and healthcare policy}
  (\bibinfo{year}{2010}) \bibinfo{pages}{61--72}.
\bibitem[{Berkman et~al.(2011)Berkman, Sheridan, Donahue, Halpern, Viera,
  Crotty, Holland, Brasure, Lohr, Harden, Tant, Wallace, and
  Viswanathan}]{berkman2011}
\bibinfo{author}{N.~D. Berkman}, \bibinfo{author}{S.~L. Sheridan},
  \bibinfo{author}{K.~E. Donahue}, \bibinfo{author}{D.~J. Halpern},
  \bibinfo{author}{A.~Viera}, \bibinfo{author}{K.~Crotty},
  \bibinfo{author}{A.~Holland}, \bibinfo{author}{M.~Brasure},
  \bibinfo{author}{K.~N. Lohr}, \bibinfo{author}{E.~Harden},
  \bibinfo{author}{E.~Tant}, \bibinfo{author}{I.~Wallace},
  \bibinfo{author}{M.~Viswanathan},
\newblock \bibinfo{title}{{Health Literacy Interventions and Outcomes: An
  Updated Systematic Review}},
\newblock \bibinfo{journal}{Evidence report/technology assessment}
  (\bibinfo{year}{2011}) \bibinfo{pages}{1—941}.
\bibitem[{Howell et~al.(2017)Howell, Harth, Brown, Bennett, and
  Boyko}]{howell2017self}
\bibinfo{author}{D.~Howell}, \bibinfo{author}{T.~Harth},
  \bibinfo{author}{J.~Brown}, \bibinfo{author}{C.~Bennett},
  \bibinfo{author}{S.~Boyko},
\newblock \bibinfo{title}{{Self-management Education Interventions for Patients
  with Cancer: A Systematic Review}},
\newblock \bibinfo{journal}{Supportive Care in Cancer} \bibinfo{volume}{25}
  (\bibinfo{year}{2017}) \bibinfo{pages}{1323--1355}.
\bibitem[{Christiansen et~al.(2023)Christiansen, Buswell, and
  Fadelu}]{christiansen2023systematic}
\bibinfo{author}{K.~Christiansen}, \bibinfo{author}{L.~Buswell},
  \bibinfo{author}{T.~Fadelu},
\newblock \bibinfo{title}{{A Systematic Review of Patient Education Strategies
  for Oncology Patients in Low-and Middle-Income Countries}},
\newblock \bibinfo{journal}{The Oncologist} \bibinfo{volume}{28}
  (\bibinfo{year}{2023}) \bibinfo{pages}{2--11}.
\bibitem[{Gilligan et~al.(2018)Gilligan, Coyle, Frankel, Berry, Bohlke,
  Epstein, Finlay, Jackson, Lathan, Loprinzi et~al.}]{gilligan2018patient}
\bibinfo{author}{T.~Gilligan}, \bibinfo{author}{N.~Coyle},
  \bibinfo{author}{R.~M. Frankel}, \bibinfo{author}{D.~L. Berry},
  \bibinfo{author}{K.~Bohlke}, \bibinfo{author}{R.~M. Epstein},
  \bibinfo{author}{E.~Finlay}, \bibinfo{author}{V.~A. Jackson},
  \bibinfo{author}{C.~S. Lathan}, \bibinfo{author}{C.~L. Loprinzi}, et~al.,
\newblock \bibinfo{title}{{Patient-clinician communication: American Society of
  Clinical Oncology Consensus Guideline}},
\newblock \bibinfo{journal}{Obstetrical \& Gynecological Survey}
  \bibinfo{volume}{73} (\bibinfo{year}{2018}) \bibinfo{pages}{96--97}.
\bibitem[{De~Groot et~al.(2019)De~Groot, Harris, Regehr, Tekian, and
  Ingledew}]{de2019quality}
\bibinfo{author}{L.~De~Groot}, \bibinfo{author}{I.~Harris},
  \bibinfo{author}{G.~Regehr}, \bibinfo{author}{A.~Tekian},
  \bibinfo{author}{P.-A. Ingledew},
\newblock \bibinfo{title}{{Quality of Online Resources for Pancreatic Cancer
  Patients}},
\newblock \bibinfo{journal}{Journal of Cancer Education} \bibinfo{volume}{34}
  (\bibinfo{year}{2019}) \bibinfo{pages}{223--228}.
\bibitem[{Alarifi et~al.(2021)Alarifi, Patrick, Jabour, Wu, and
  Luo}]{alarifi2021understanding}
\bibinfo{author}{M.~Alarifi}, \bibinfo{author}{T.~Patrick},
  \bibinfo{author}{A.~Jabour}, \bibinfo{author}{M.~Wu},
  \bibinfo{author}{J.~Luo},
\newblock \bibinfo{title}{{Understanding {P}atient {N}eeds and {G}aps in
  {R}adiology {R}eports {T}hrough {O}nline {D}iscussion {F}orum {A}nalysis}},
\newblock \bibinfo{journal}{Insights into Imaging} \bibinfo{volume}{12}
  (\bibinfo{year}{2021}) \bibinfo{pages}{1--9}.
\bibitem[{Yang et~al.(2023)Yang, Li, Lin, Wang, Lin, Liu, and
  Wang}]{yang2023dawn}
\bibinfo{author}{Z.~Yang}, \bibinfo{author}{L.~Li}, \bibinfo{author}{K.~Lin},
  \bibinfo{author}{J.~Wang}, \bibinfo{author}{C.-C. Lin},
  \bibinfo{author}{Z.~Liu}, \bibinfo{author}{L.~Wang},
\newblock \bibinfo{title}{The {D}awn of {LMMs}: {P}reliminary {E}xplorations
  with {GPT-4V}(ision)},
\newblock \bibinfo{journal}{arXiv preprint arXiv:2309.17421}
  \bibinfo{volume}{9} (\bibinfo{year}{2023}).
\bibitem[{Chowdhery et~al.(2023)Chowdhery, Narang, Devlin, Bosma, Mishra,
  Roberts, Barham, Chung, Sutton, Gehrmann et~al.}]{chowdhery2023palm}
\bibinfo{author}{A.~Chowdhery}, \bibinfo{author}{S.~Narang},
  \bibinfo{author}{J.~Devlin}, \bibinfo{author}{M.~Bosma},
  \bibinfo{author}{G.~Mishra}, \bibinfo{author}{A.~Roberts},
  \bibinfo{author}{P.~Barham}, \bibinfo{author}{H.~W. Chung},
  \bibinfo{author}{C.~Sutton}, \bibinfo{author}{S.~Gehrmann}, et~al.,
\newblock \bibinfo{title}{Palm: {S}caling {L}anguage {M}odeling with
  {P}athways},
\newblock \bibinfo{journal}{Journal of Machine Learning Research}
  \bibinfo{volume}{24} (\bibinfo{year}{2023}) \bibinfo{pages}{1--113}.
\bibitem[{Radford et~al.(2019)Radford, Wu, Child, Luan, Amodei, Sutskever
  et~al.}]{radford2019language}
\bibinfo{author}{A.~Radford}, \bibinfo{author}{J.~Wu},
  \bibinfo{author}{R.~Child}, \bibinfo{author}{D.~Luan},
  \bibinfo{author}{D.~Amodei}, \bibinfo{author}{I.~Sutskever}, et~al.,
\newblock \bibinfo{title}{Language {M}odels are {U}nsupervised {M}ultitask
  {L}earners},
\newblock \bibinfo{journal}{OpenAI blog} \bibinfo{volume}{1}
  (\bibinfo{year}{2019}) \bibinfo{pages}{9}.
\bibitem[{Brown et~al.(2020)Brown, Mann, Ryder, Subbiah, Kaplan, Dhariwal,
  Neelakantan, Shyam, Sastry, Askell et~al.}]{brown2020language}
\bibinfo{author}{T.~Brown}, \bibinfo{author}{B.~Mann},
  \bibinfo{author}{N.~Ryder}, \bibinfo{author}{M.~Subbiah},
  \bibinfo{author}{J.~D. Kaplan}, \bibinfo{author}{P.~Dhariwal},
  \bibinfo{author}{A.~Neelakantan}, \bibinfo{author}{P.~Shyam},
  \bibinfo{author}{G.~Sastry}, \bibinfo{author}{A.~Askell}, et~al.,
\newblock \bibinfo{title}{Language {M}odels are {F}ew-shot {L}earners},
\newblock \bibinfo{journal}{Advances in neural information processing systems}
  \bibinfo{volume}{33} (\bibinfo{year}{2020}) \bibinfo{pages}{1877--1901}.
\bibitem[{OpenAI et~al.(2023)OpenAI, :, Achiam, Adler, Agarwal, Ahmad, Akkaya,
  Aleman, Almeida, Altenschmidt, Altman, Anadkat, Avila, Babuschkin, Balaji,
  Balcom, Baltescu et~al.}]{openai2023gpt4}
\bibinfo{author}{OpenAI}, \bibinfo{author}{:}, \bibinfo{author}{J.~Achiam},
  \bibinfo{author}{S.~Adler}, \bibinfo{author}{S.~Agarwal},
  \bibinfo{author}{L.~Ahmad}, \bibinfo{author}{I.~Akkaya},
  \bibinfo{author}{F.~L. Aleman}, \bibinfo{author}{D.~Almeida},
  \bibinfo{author}{J.~Altenschmidt}, \bibinfo{author}{S.~Altman},
  \bibinfo{author}{S.~Anadkat}, \bibinfo{author}{R.~Avila},
  \bibinfo{author}{I.~Babuschkin}, \bibinfo{author}{S.~Balaji},
  \bibinfo{author}{V.~Balcom}, \bibinfo{author}{P.~Baltescu}, et~al.,
  \bibinfo{title}{Gpt-4 {T}echnical {R}eport}, \bibinfo{year}{2023}.
  \href{http://arxiv.org/abs/2303.08774}{{\tt arXiv:2303.08774}}.
\bibitem[{Touvron et~al.(2023{\natexlab{a}})Touvron, Lavril, Izacard, Martinet,
  Lachaux, Lacroix, Rozi{\`e}re, Goyal, Hambro, Azhar
  et~al.}]{touvron2023llama}
\bibinfo{author}{H.~Touvron}, \bibinfo{author}{T.~Lavril},
  \bibinfo{author}{G.~Izacard}, \bibinfo{author}{X.~Martinet},
  \bibinfo{author}{M.-A. Lachaux}, \bibinfo{author}{T.~Lacroix},
  \bibinfo{author}{B.~Rozi{\`e}re}, \bibinfo{author}{N.~Goyal},
  \bibinfo{author}{E.~Hambro}, \bibinfo{author}{F.~Azhar}, et~al.,
\newblock \bibinfo{title}{Llama: {O}pen and {E}fficient {F}oundation {L}anguage
  {M}odels},
\newblock \bibinfo{journal}{arXiv preprint arXiv:2302.13971}
  (\bibinfo{year}{2023}{\natexlab{a}}).
\bibitem[{Touvron et~al.(2023{\natexlab{b}})Touvron, Martin, Stone, Albert,
  Almahairi, Babaei, Bashlykov, Batra, Bhargava, Bhosale
  et~al.}]{touvron2023llama2}
\bibinfo{author}{H.~Touvron}, \bibinfo{author}{L.~Martin},
  \bibinfo{author}{K.~Stone}, \bibinfo{author}{P.~Albert},
  \bibinfo{author}{A.~Almahairi}, \bibinfo{author}{Y.~Babaei},
  \bibinfo{author}{N.~Bashlykov}, \bibinfo{author}{S.~Batra},
  \bibinfo{author}{P.~Bhargava}, \bibinfo{author}{S.~Bhosale}, et~al.,
\newblock \bibinfo{title}{Llama 2: {O}pen {F}oundation and {F}ine-tuned {C}hat
  {M}odels},
\newblock \bibinfo{journal}{arXiv preprint arXiv:2307.09288}
  (\bibinfo{year}{2023}{\natexlab{b}}).
\bibitem[{Anthropic(2024)}]{claude2024}
\bibinfo{author}{Anthropic}, \bibinfo{title}{Claude 3.5 sonnet},
  \bibinfo{year}{2024}. \URLprefix
  \url{https://www.anthropic.com/news/claude-3-5-sonnet}.
\bibitem[{Zhao et~al.(2023)Zhao, Zhou, Li, Tang, Wang, Hou, Min, Zhang, Zhang,
  Dong et~al.}]{zhao2023survey}
\bibinfo{author}{W.~X. Zhao}, \bibinfo{author}{K.~Zhou},
  \bibinfo{author}{J.~Li}, \bibinfo{author}{T.~Tang},
  \bibinfo{author}{X.~Wang}, \bibinfo{author}{Y.~Hou},
  \bibinfo{author}{Y.~Min}, \bibinfo{author}{B.~Zhang},
  \bibinfo{author}{J.~Zhang}, \bibinfo{author}{Z.~Dong}, et~al.,
\newblock \bibinfo{title}{A {S}urvey of {L}arge {L}anguage {M}odels},
\newblock \bibinfo{journal}{arXiv preprint arXiv:2303.18223}
  (\bibinfo{year}{2023}).
\bibitem[{Yang et~al.(2023)Yang, Jin, Tang, Han, Feng, Jiang, Yin, and
  Hu}]{yang2023harnessing}
\bibinfo{author}{J.~Yang}, \bibinfo{author}{H.~Jin}, \bibinfo{author}{R.~Tang},
  \bibinfo{author}{X.~Han}, \bibinfo{author}{Q.~Feng},
  \bibinfo{author}{H.~Jiang}, \bibinfo{author}{B.~Yin},
  \bibinfo{author}{X.~Hu},
\newblock \bibinfo{title}{Harnessing the {P}ower of {LLMs} in {P}ractice: A
  {S}urvey on {C}hatgpt and {B}eyond},
\newblock \bibinfo{journal}{arXiv preprint arXiv:2304.13712}
  (\bibinfo{year}{2023}).
\bibitem[{Singhal et~al.(2023{\natexlab{a}})Singhal, Azizi, Tu, Mahdavi, Wei,
  Chung, Scales, Tanwani, Cole-Lewis, Pfohl et~al.}]{singhal2023large}
\bibinfo{author}{K.~Singhal}, \bibinfo{author}{S.~Azizi},
  \bibinfo{author}{T.~Tu}, \bibinfo{author}{S.~S. Mahdavi},
  \bibinfo{author}{J.~Wei}, \bibinfo{author}{H.~W. Chung},
  \bibinfo{author}{N.~Scales}, \bibinfo{author}{A.~Tanwani},
  \bibinfo{author}{H.~Cole-Lewis}, \bibinfo{author}{S.~Pfohl}, et~al.,
\newblock \bibinfo{title}{Large {L}anguage {M}odels {E}ncode {C}linical
  {K}nowledge},
\newblock \bibinfo{journal}{Nature} \bibinfo{volume}{620}
  (\bibinfo{year}{2023}{\natexlab{a}}) \bibinfo{pages}{172--180}.
\bibitem[{Singhal et~al.(2023{\natexlab{b}})Singhal, Tu, Gottweis, Sayres,
  Wulczyn, Hou, Clark, Pfohl, Cole-Lewis, Neal et~al.}]{singhal2023towards}
\bibinfo{author}{K.~Singhal}, \bibinfo{author}{T.~Tu},
  \bibinfo{author}{J.~Gottweis}, \bibinfo{author}{R.~Sayres},
  \bibinfo{author}{E.~Wulczyn}, \bibinfo{author}{L.~Hou},
  \bibinfo{author}{K.~Clark}, \bibinfo{author}{S.~Pfohl},
  \bibinfo{author}{H.~Cole-Lewis}, \bibinfo{author}{D.~Neal}, et~al.,
\newblock \bibinfo{title}{Towards {E}xpert-level {M}edical {Q}uestion
  {A}nswering with {L}arge {L}anguage {M}odels},
\newblock \bibinfo{journal}{arXiv preprint arXiv:2305.09617}
  (\bibinfo{year}{2023}{\natexlab{b}}).
\bibitem[{Han et~al.(2023)Han, Adams, Papaioannou, Grundmann, Oberhauser,
  L{\"o}ser, Truhn, and Bressem}]{han2023medalpaca}
\bibinfo{author}{T.~Han}, \bibinfo{author}{L.~C. Adams}, \bibinfo{author}{J.-M.
  Papaioannou}, \bibinfo{author}{P.~Grundmann},
  \bibinfo{author}{T.~Oberhauser}, \bibinfo{author}{A.~L{\"o}ser},
  \bibinfo{author}{D.~Truhn}, \bibinfo{author}{K.~K. Bressem},
\newblock \bibinfo{title}{{MedAlpaca}: {A}n {O}pen-{S}ource {C}ollection of
  {M}edical {C}onversational {AI} {M}odels and {T}raining {D}ata},
\newblock \bibinfo{journal}{arXiv preprint arXiv:2304.08247}
  (\bibinfo{year}{2023}).
\bibitem[{Toma et~al.(2023)Toma, Lawler, Ba, Krishnan, Rubin, and
  Wang}]{toma2023clinical}
\bibinfo{author}{A.~Toma}, \bibinfo{author}{P.~R. Lawler},
  \bibinfo{author}{J.~Ba}, \bibinfo{author}{R.~G. Krishnan},
  \bibinfo{author}{B.~B. Rubin}, \bibinfo{author}{B.~Wang},
\newblock \bibinfo{title}{Clinical {C}amel: {A}n {O}pen-{S}ource {E}xpert-level
  {M}edical {L}anguage {M}odel with {D}ialogue-based {K}nowledge {E}ncoding},
\newblock \bibinfo{journal}{arXiv preprint arXiv:2305.12031}
  (\bibinfo{year}{2023}).
\bibitem[{Yunxiang et~al.(2023)Yunxiang, Zihan, Kai, Ruilong, and
  You}]{yunxiang2023chatdoctor}
\bibinfo{author}{L.~Yunxiang}, \bibinfo{author}{L.~Zihan},
  \bibinfo{author}{Z.~Kai}, \bibinfo{author}{D.~Ruilong},
  \bibinfo{author}{Z.~You},
\newblock \bibinfo{title}{Chatdoctor: {A} {M}edical {C}hat {M}odel {F}ine-tuned
  on llama {M}odel using {M}edical {D}omain {K}nowledge},
\newblock \bibinfo{journal}{arXiv preprint arXiv:2303.14070}
  (\bibinfo{year}{2023}).
\bibitem[{Liu et~al.(2023)Liu, Zhu, Wu, Yang, You, Wang, Lu, Liu, Zheng, Sun
  et~al.}]{liu2023medical}
\bibinfo{author}{F.~Liu}, \bibinfo{author}{T.~Zhu}, \bibinfo{author}{X.~Wu},
  \bibinfo{author}{B.~Yang}, \bibinfo{author}{C.~You},
  \bibinfo{author}{C.~Wang}, \bibinfo{author}{L.~Lu}, \bibinfo{author}{Z.~Liu},
  \bibinfo{author}{Y.~Zheng}, \bibinfo{author}{X.~Sun}, et~al.,
\newblock \bibinfo{title}{A {M}edical {M}ultimodal {L}arge {L}anguage {M}odel
  for {F}uture {P}andemics},
\newblock \bibinfo{journal}{NPJ Digital Medicine} \bibinfo{volume}{6}
  (\bibinfo{year}{2023}) \bibinfo{pages}{226}.
\bibitem[{Arora and Arora(2023)}]{arora2023promise}
\bibinfo{author}{A.~Arora}, \bibinfo{author}{A.~Arora},
\newblock \bibinfo{title}{The {P}romise of {L}arge {L}anguage {M}odels in
  {H}ealthcare},
\newblock \bibinfo{journal}{The Lancet} \bibinfo{volume}{401}
  (\bibinfo{year}{2023}) \bibinfo{pages}{641}.
\bibitem[{Thirunavukarasu et~al.(2023)Thirunavukarasu, Ting, Elangovan,
  Gutierrez, Tan, and Ting}]{thirunavukarasu2023large}
\bibinfo{author}{A.~J. Thirunavukarasu}, \bibinfo{author}{D.~S.~J. Ting},
  \bibinfo{author}{K.~Elangovan}, \bibinfo{author}{L.~Gutierrez},
  \bibinfo{author}{T.~F. Tan}, \bibinfo{author}{D.~S.~W. Ting},
\newblock \bibinfo{title}{Large {L}anguage {M}odels in {M}edicine},
\newblock \bibinfo{journal}{Nature medicine} \bibinfo{volume}{29}
  (\bibinfo{year}{2023}) \bibinfo{pages}{1930--1940}.
\bibitem[{Patel and Lam(2023)}]{patel2023chatgpt}
\bibinfo{author}{S.~B. Patel}, \bibinfo{author}{K.~Lam},
\newblock \bibinfo{title}{{ChatGPT}: {T}he {F}uture of {D}ischarge
  {S}ummaries?},
\newblock \bibinfo{journal}{The Lancet Digital Health} \bibinfo{volume}{5}
  (\bibinfo{year}{2023}) \bibinfo{pages}{e107--e108}.
\bibitem[{Wu et~al.(2023)Wu, Zhang, Zhang, Wang, and Xie}]{wu2023pmc}
\bibinfo{author}{C.~Wu}, \bibinfo{author}{X.~Zhang},
  \bibinfo{author}{Y.~Zhang}, \bibinfo{author}{Y.~Wang},
  \bibinfo{author}{W.~Xie},
\newblock \bibinfo{title}{Pmc-llama: {F}urther {F}inetuning llama on {M}edical
  {P}apers},
\newblock \bibinfo{journal}{arXiv preprint arXiv:2304.14454}
  (\bibinfo{year}{2023}).
\bibitem[{Wang et~al.(2023)Wang, Liu, Xi, Qiang, Zhao, Qin, and
  Liu}]{wang2023huatuo}
\bibinfo{author}{H.~Wang}, \bibinfo{author}{C.~Liu}, \bibinfo{author}{N.~Xi},
  \bibinfo{author}{Z.~Qiang}, \bibinfo{author}{S.~Zhao},
  \bibinfo{author}{B.~Qin}, \bibinfo{author}{T.~Liu},
\newblock \bibinfo{title}{Huatuo: {T}uning llama {M}odel with {C}hinese
  {M}edical {K}nowledge},
\newblock \bibinfo{journal}{arXiv preprint arXiv:2304.06975}
  (\bibinfo{year}{2023}).
\bibitem[{Liu et~al.(2023)Liu, Li, Wu, and Lee}]{liu2023visual}
\bibinfo{author}{H.~Liu}, \bibinfo{author}{C.~Li}, \bibinfo{author}{Q.~Wu},
  \bibinfo{author}{Y.~J. Lee},
\newblock \bibinfo{title}{Visual {I}nstruction {T}uning},
\newblock \bibinfo{journal}{arXiv preprint arXiv:2304.08485}
  (\bibinfo{year}{2023}).
\bibitem[{Awadalla et~al.(2023)Awadalla, Gao, Gardner, Hessel, Hanafy, Zhu,
  Marathe, Bitton, Gadre, Sagawa et~al.}]{awadalla2023openflamingo}
\bibinfo{author}{A.~Awadalla}, \bibinfo{author}{I.~Gao},
  \bibinfo{author}{J.~Gardner}, \bibinfo{author}{J.~Hessel},
  \bibinfo{author}{Y.~Hanafy}, \bibinfo{author}{W.~Zhu},
  \bibinfo{author}{K.~Marathe}, \bibinfo{author}{Y.~Bitton},
  \bibinfo{author}{S.~Gadre}, \bibinfo{author}{S.~Sagawa}, et~al.,
\newblock \bibinfo{title}{Openflamingo: {A}n {O}pen-source {F}ramework for
  {T}raining {L}arge {A}utoregressive {V}ision-language {M}odels},
\newblock \bibinfo{journal}{arXiv preprint arXiv:2308.01390}
  (\bibinfo{year}{2023}).
\bibitem[{Li et~al.(2023)Li, Li, Savarese, and Hoi}]{li2023blip}
\bibinfo{author}{J.~Li}, \bibinfo{author}{D.~Li},
  \bibinfo{author}{S.~Savarese}, \bibinfo{author}{S.~Hoi},
\newblock \bibinfo{title}{Blip-2: {B}ootstrapping {L}anguage-image
  {P}re-training with {F}rozen {I}mage {E}ncoders and {L}arge {L}anguage
  {M}odels},
\newblock \bibinfo{journal}{arXiv preprint arXiv:2301.12597}
  (\bibinfo{year}{2023}).
\bibitem[{Liu et~al.(2023)Liu, Hyland, Bannur, Bouzid, Castro, Wetscherek,
  Tinn, Sharma, P{\'e}rez-Garc{\'\i}a, Schwaighofer et~al.}]{liu2023exploring}
\bibinfo{author}{Q.~Liu}, \bibinfo{author}{S.~Hyland},
  \bibinfo{author}{S.~Bannur}, \bibinfo{author}{K.~Bouzid},
  \bibinfo{author}{D.~C. Castro}, \bibinfo{author}{M.~T. Wetscherek},
  \bibinfo{author}{R.~Tinn}, \bibinfo{author}{H.~Sharma},
  \bibinfo{author}{F.~P{\'e}rez-Garc{\'\i}a},
  \bibinfo{author}{A.~Schwaighofer}, et~al.,
\newblock \bibinfo{title}{Exploring the {B}oundaries of {GPT-4} in
  {R}adiology},
\newblock \bibinfo{journal}{arXiv preprint arXiv:2310.14573}
  (\bibinfo{year}{2023}).
\bibitem[{Bannur et~al.(2023)Bannur, Hyland, Liu, P{\'e}rez-Garc{\'\i}a, Ilse,
  de~Castro, Boecking, Sharma, Bouzid, Schwaighofer et~al.}]{bannur2023mscxr}
\bibinfo{author}{S.~Bannur}, \bibinfo{author}{S.~Hyland},
  \bibinfo{author}{Q.~Liu}, \bibinfo{author}{F.~P{\'e}rez-Garc{\'\i}a},
  \bibinfo{author}{M.~Ilse}, \bibinfo{author}{D.~C. de~Castro},
  \bibinfo{author}{B.~Boecking}, \bibinfo{author}{H.~Sharma},
  \bibinfo{author}{K.~Bouzid}, \bibinfo{author}{A.~Schwaighofer}, et~al.,
  \bibinfo{title}{{MSCXR-T}: {L}earning to {E}xploit {T}emporal {S}tructure for
  {B}iomedical {V}ision-{L}anguage {P}rocessing}, \bibinfo{year}{2023}.
\bibitem[{Smit et~al.(2020)Smit, Jain, Rajpurkar, Pareek, Ng, and
  Lungren}]{smit2020chexbert}
\bibinfo{author}{A.~Smit}, \bibinfo{author}{S.~Jain},
  \bibinfo{author}{P.~Rajpurkar}, \bibinfo{author}{A.~Pareek},
  \bibinfo{author}{A.~Y. Ng}, \bibinfo{author}{M.~P. Lungren},
\newblock \bibinfo{title}{{CheXbert}: {C}ombining {A}utomatic {L}abelers and
  {E}xpert {A}nnotations for {A}ccurate {R}adiology {R}eport {L}abeling using
  {BERT}},
\newblock \bibinfo{journal}{arXiv preprint arXiv:2004.09167}
  (\bibinfo{year}{2020}).
\bibitem[{Xie et~al.(2024)Xie, Palayew, Toma, Bader, and
  Wang}]{toma-etal-2024-wanglab-mediqa}
\bibinfo{author}{R.~Xie}, \bibinfo{author}{S.~Palayew},
  \bibinfo{author}{A.~Toma}, \bibinfo{author}{G.~Bader},
  \bibinfo{author}{B.~Wang},
\newblock \bibinfo{title}{{W}ang{L}ab at {MEDIQA}-{M}3{G} 2024: Multimodal
  medical answer generation using large language models},
\newblock in: \bibinfo{editor}{T.~Naumann}, \bibinfo{editor}{A.~Ben~Abacha},
  \bibinfo{editor}{S.~Bethard}, \bibinfo{editor}{K.~Roberts},
  \bibinfo{editor}{D.~Bitterman} (Eds.), \bibinfo{booktitle}{Proceedings of the
  6th Clinical Natural Language Processing Workshop},
  \bibinfo{publisher}{Association for Computational Linguistics},
  \bibinfo{address}{Mexico City, Mexico}, \bibinfo{year}{2024}, pp.
  \bibinfo{pages}{624--634}. \URLprefix
  \url{https://aclanthology.org/2024.clinicalnlp-1.60}.
  \DOIprefix\doi{10.18653/v1/2024.clinicalnlp-1.60}.
\bibitem[{He et~al.(2023)He, Mao, Lin, Ruan, Lan, Feng, and
  Cambria}]{he2023survey}
\bibinfo{author}{K.~He}, \bibinfo{author}{R.~Mao}, \bibinfo{author}{Q.~Lin},
  \bibinfo{author}{Y.~Ruan}, \bibinfo{author}{X.~Lan},
  \bibinfo{author}{M.~Feng}, \bibinfo{author}{E.~Cambria},
\newblock \bibinfo{title}{A {S}urvey of {L}arge {L}anguage {M}odels for
  {H}ealthcare: {F}rom {D}ata, {T}echnology, and {A}pplications to
  {A}ccountability and {E}thics},
\newblock \bibinfo{journal}{arXiv preprint arXiv:2310.05694}
  (\bibinfo{year}{2023}).
\bibitem[{Xie et~al.(2023)Xie, Schenck, Yang, Chen, Peng, and
  Wang}]{xie2023faithful}
\bibinfo{author}{Q.~Xie}, \bibinfo{author}{E.~J. Schenck},
  \bibinfo{author}{H.~S. Yang}, \bibinfo{author}{Y.~Chen},
  \bibinfo{author}{Y.~Peng}, \bibinfo{author}{F.~Wang},
\newblock \bibinfo{title}{Faithful {AI} in {M}edicine: {A} {S}ystematic
  {R}eview with {L}arge {L}anguage {M}odels and {B}eyond},
\newblock \bibinfo{journal}{medRxiv}  (\bibinfo{year}{2023}).
\bibitem[{Pal et~al.(2022)Pal, Umapathi, and Sankarasubbu}]{pal2022medmcqa}
\bibinfo{author}{A.~Pal}, \bibinfo{author}{L.~K. Umapathi},
  \bibinfo{author}{M.~Sankarasubbu},
\newblock \bibinfo{title}{Medmcqa: {A} {L}arge-scale {M}ulti-subject
  {M}ulti-choice {D}ataset for {M}edical {D}omain {Q}uestion {A}nswering},
\newblock in: \bibinfo{booktitle}{Conference on Health, Inference, and
  Learning}, \bibinfo{organization}{PMLR}, \bibinfo{year}{2022}, pp.
  \bibinfo{pages}{248--260}.
\bibitem[{Lin et~al.(2021)Lin, Hilton, and Evans}]{lin2021truthfulqa}
\bibinfo{author}{S.~Lin}, \bibinfo{author}{J.~Hilton},
  \bibinfo{author}{O.~Evans},
\newblock \bibinfo{title}{Truthfulqa: {M}easuring {H}ow {M}odels {M}imic
  {H}uman {F}alsehoods},
\newblock \bibinfo{journal}{arXiv preprint arXiv:2109.07958}
  (\bibinfo{year}{2021}).
\bibitem[{Li et~al.(2023)Li, Cheng, Zhao, Nie, and Wen}]{li2023halueval}
\bibinfo{author}{J.~Li}, \bibinfo{author}{X.~Cheng}, \bibinfo{author}{W.~X.
  Zhao}, \bibinfo{author}{J.-Y. Nie}, \bibinfo{author}{J.-R. Wen},
\newblock \bibinfo{title}{Halueval: {A} {L}arge-scale {H}allucination
  {E}valuation {B}enchmark for {L}arge {L}anguage {M}odels},
\newblock in: \bibinfo{booktitle}{Proceedings of the 2023 Conference on
  Empirical Methods in Natural Language Processing}, \bibinfo{year}{2023}, pp.
  \bibinfo{pages}{6449--6464}.
\bibitem[{Glazer and Ruiz-Wibbelsmann(2011)}]{glazer2011invisible}
\bibinfo{author}{G.~M. Glazer}, \bibinfo{author}{J.~A. Ruiz-Wibbelsmann},
\newblock \bibinfo{title}{{The Invisible Radiologist}},
\newblock \bibinfo{journal}{Radiology} \bibinfo{volume}{258}
  (\bibinfo{year}{2011}) \bibinfo{pages}{18--22}.
\bibitem[{Kemp et~al.(2017)Kemp, Mahoney, Mathews, Wintermark, Yee, and
  Brown}]{kemp2017patient}
\bibinfo{author}{J.~L. Kemp}, \bibinfo{author}{M.~C. Mahoney},
  \bibinfo{author}{V.~P. Mathews}, \bibinfo{author}{M.~Wintermark},
  \bibinfo{author}{J.~Yee}, \bibinfo{author}{S.~D. Brown},
\newblock \bibinfo{title}{{Patient-Centered Radiology: Where are We, Where do
  We Want to Be, and How do We Get There?}},
\newblock \bibinfo{journal}{Radiology} \bibinfo{volume}{285}
  (\bibinfo{year}{2017}) \bibinfo{pages}{601--608}.
\bibitem[{Nowell et~al.(2017)Nowell, Norris, White, and
  Moules}]{nowell2017thematic}
\bibinfo{author}{L.~S. Nowell}, \bibinfo{author}{J.~M. Norris},
  \bibinfo{author}{D.~E. White}, \bibinfo{author}{N.~J. Moules},
\newblock \bibinfo{title}{{Thematic Analysis: Striving to Meet the
  Trustworthiness Criteria}},
\newblock \bibinfo{journal}{International journal of qualitative methods}
  \bibinfo{volume}{16} (\bibinfo{year}{2017})
  \bibinfo{pages}{1609406917733847}.
\bibitem[{OpenAI(2024)}]{gpt2024}
\bibinfo{author}{OpenAI}, \bibinfo{title}{Hello gpt-4o}, \bibinfo{year}{2024}.
  \URLprefix \url{https://openai.com/index/hello-gpt-4o/}.
\bibitem[{Wang et~al.(2024)Wang, Ma, Zhang, Ni, Chandra, Guo, Ren, Arulraj, He,
  Jiang et~al.}]{wang2024mmlu}
\bibinfo{author}{Y.~Wang}, \bibinfo{author}{X.~Ma}, \bibinfo{author}{G.~Zhang},
  \bibinfo{author}{Y.~Ni}, \bibinfo{author}{A.~Chandra},
  \bibinfo{author}{S.~Guo}, \bibinfo{author}{W.~Ren},
  \bibinfo{author}{A.~Arulraj}, \bibinfo{author}{X.~He},
  \bibinfo{author}{Z.~Jiang}, et~al.,
\newblock \bibinfo{title}{Mmlu-pro: A more robust and challenging multi-task
  language understanding benchmark},
\newblock \bibinfo{journal}{arXiv preprint arXiv:2406.01574}
  (\bibinfo{year}{2024}).
\bibitem[{Lin(2004)}]{lin2004rouge}
\bibinfo{author}{C.-Y. Lin},
\newblock \bibinfo{title}{{ROUGE: A Package for Automatic Evaluation of
  Summaries}},
\newblock in: \bibinfo{booktitle}{Text Summarization Branches Out, Proceedings
  of the ACL-04 Workshop}, \bibinfo{year}{2004}, pp. \bibinfo{pages}{74--81}.
\bibitem[{Cer et~al.(2018)Cer, Yang, Kong, Hua, Limtiaco, John, Constant,
  Guajardo-Cespedes, Yuan, Tar et~al.}]{cer2018universal}
\bibinfo{author}{D.~Cer}, \bibinfo{author}{Y.~Yang}, \bibinfo{author}{S.-y.
  Kong}, \bibinfo{author}{N.~Hua}, \bibinfo{author}{N.~Limtiaco},
  \bibinfo{author}{R.~S. John}, \bibinfo{author}{N.~Constant},
  \bibinfo{author}{M.~Guajardo-Cespedes}, \bibinfo{author}{S.~Yuan},
  \bibinfo{author}{C.~Tar}, et~al.,
\newblock \bibinfo{title}{Universal sentence encoder for english},
\newblock in: \bibinfo{booktitle}{Proceedings of the 2018 conference on
  empirical methods in natural language processing: system demonstrations},
  \bibinfo{year}{2018}, pp. \bibinfo{pages}{169--174}.
\bibitem[{Grice(1975)}]{grice1975logic}
\bibinfo{author}{H.~Grice},
\newblock \bibinfo{title}{Logic and conversation},
\newblock \bibinfo{journal}{Syntax and semantics} \bibinfo{volume}{3}
  (\bibinfo{year}{1975}).
\bibitem[{Bubeck et~al.(2023)Bubeck, Chandrasekaran, Eldan, Gehrke, Horvitz,
  Kamar, Lee, Lee, Li, Lundberg et~al.}]{bubeck2023sparks}
\bibinfo{author}{S.~Bubeck}, \bibinfo{author}{V.~Chandrasekaran},
  \bibinfo{author}{R.~Eldan}, \bibinfo{author}{J.~Gehrke},
  \bibinfo{author}{E.~Horvitz}, \bibinfo{author}{E.~Kamar},
  \bibinfo{author}{P.~Lee}, \bibinfo{author}{Y.~T. Lee},
  \bibinfo{author}{Y.~Li}, \bibinfo{author}{S.~Lundberg}, et~al.,
\newblock \bibinfo{title}{Sparks of artificial general intelligence: Early
  experiments with gpt-4},
\newblock \bibinfo{journal}{arXiv preprint arXiv:2303.12712}
  (\bibinfo{year}{2023}).
\bibitem[{Papineni et~al.(2002)Papineni, Roukos, Ward, and
  Zhu}]{papineni2002bleu}
\bibinfo{author}{K.~Papineni}, \bibinfo{author}{S.~Roukos},
  \bibinfo{author}{T.~Ward}, \bibinfo{author}{W.-J. Zhu},
\newblock \bibinfo{title}{{BLEU: A Method for Automatic Evaluation of Machine
  Translation}},
\newblock in: \bibinfo{booktitle}{Proceedings of the 40th Annual Meeting of the
  Association for Computational Linguistics},
  \bibinfo{organization}{Association for Computational Linguistics},
  \bibinfo{year}{2002}, pp. \bibinfo{pages}{311--318}.
\bibitem[{Zhang et~al.(2020)Zhang, Kishore, Wu, Weinberger, and
  Artzi}]{zhang2020bertscore}
\bibinfo{author}{T.~Zhang}, \bibinfo{author}{V.~Kishore},
  \bibinfo{author}{F.~Wu}, \bibinfo{author}{K.~Q. Weinberger},
  \bibinfo{author}{Y.~Artzi},
\newblock \bibinfo{title}{{BERTScore: Evaluating Text Generation with BERT}},
\newblock in: \bibinfo{booktitle}{International Conference on Learning
  Representations}, \bibinfo{year}{2020}.

\end{thebibliography}
\end{document}